\documentclass{emulateapj} 

\usepackage{epsfig,color}
\usepackage{times}
\usepackage{latexsym}
\usepackage{url}

\usepackage{natbib}
\bibpunct{(}{)}{;}{a}{}{,}
\def\spose#1{\hbox to 0pt{#1\hss}}
\def\ltsimm{\mathrel{\spose{\lower 3pt\hbox{$\sim$}}
        \raise 2.0pt\hbox{$<$}}}
\def\gtsimm{\mathrel{\spose{\lower 3pt\hbox{$\sim$}}
        \raise 2.0pt\hbox{$>$}}}

\def\cm{{\rm\thinspace cm}}

\def\s{{\rm\thinspace s}}
\def\yr{{\rm\thinspace yr}}
\def\g{{\rm\thinspace g}}

\def\erg{{\rm\thinspace erg}}
\def\Hz{{\rm\thinspace Hz}}

\def\ster{{\rm\thinspace ster}}
\def\ergps{\hbox{${\rm\erg\s^{-1}\,}$}}
\def\Rsol{\hbox{${\rm\thinspace R_{\odot}}$}}
\def\Msol{\hbox{${\rm\thinspace M_{\odot}}$}}
\def\Lsol{\hbox{${\rm\thinspace L_{\odot}}$}}
\def\Msolpyr{\hbox{${\rm\Msol\yr^{-1}\,}$}}
\def\pcm{\hbox{${\rm\cm^{-1}\,}$}}
\def\pcm3{\hbox{${\rm\cm^{-3}\,}$}}
\def\ergpscm3Hz{\hbox{${\rm\ergps\cm^{-3}\Hz^{-1}\,}$}}
\def\ergpscm3Hzster{\hbox{${\rm\ergps\cm^{-3}\Hz^{-1}\ster^{-1}\,}$}}
\def\gpcm3{\hbox{${\rm\g\cm^{-3}\,}$}}
\def\ergpcm2{\hbox{${\rm\erg\cm^{-2}\,}$}}
\def\ergpcm3{\hbox{${\rm\erg\cm^{-3}\,}$}}
\def\phpscm2{\hbox{${\rm photons\s^{-1}\cm^{-2}\,}$}}

\def\pcm2{\hbox{${\rm\cm^{-2}\,}$}}
\newcommand{\anchorfoot}[2] {\anchor{#1}{#2}\footnote{\url{#1}}}
\newcommand{\anchorparen}[2]{\anchor{#1}{#2} (\url{#1})}
\def\aap{{\rm A\&A}}
\def\aas{{\rm A\&AS}}
\def\apj{{\rm ApJ}}
\def\apjs{{\rm ApJS}}
\def\aj{{\rm AJ}}
\def\mnras{{\rm MNRAS}}

\value{1}

\begin{document}

\title{X-ray emission from the double-binary OB-star system QZ Car (HD
  93206)}

\shorttitle{X-ray emission from QZ Car}

\author{E.~R.~Parkin\altaffilmark{1,2}, P.~S.~Broos\altaffilmark{3},
  L.~K.~Townsley\altaffilmark{3},  J.~M.~Pittard\altaffilmark{2},
  A.~F.~J.~Moffat\altaffilmark{4}, 
  Y.~Naz\'{e}\altaffilmark{1}, G.~Rauw\altaffilmark{1}, 
  L.~M.~Oskinova\altaffilmark{5}, W.~L.~Waldron\altaffilmark{6}}
\affil{$^{1}$ Institut d'Astrophysique et de G\'{e}ophysique,
  Universit\'{e} de Li\`{e}ge, 17, All\'{e}e du 6 Ao\^{u}t, B5c,
  B-4000 Sart Tilman, Belgium}
\affil{$^{2}$ School of Physics and Astronomy, The University of Leeds,
  Woodhouse Lane, Leeds LS2 9JT, UK} \email{email: parkin@mso.anu.edu.au}
\affil{$^{3}$ Department of Astronomy and Astrophysics, Pennsylvania
  State University, 525 Davey Laboratory, University Park, PA 16802}
\affil{$^{4}$ D\'{e}partement de Physique, Universit\'{e} de Montreal, C.P. 6128, Succ. Centre-Ville, Montreal, QC H3C 3J7, Canada}
\affil{$^{5}$ Institute for Physics and Astronomy, University of Potsdam, 14476 Potsdam, Germany}
\affil{$^{6}$ Eureka Scientific Inc., 2452 Delmer Street, Oakland, CA 94602, USA}
\shortauthors{E.~R.~Parkin et al.}


\label{firstpage}

\begin{abstract}
X-ray observations of the double-binary OB-star system QZ~Car
(HD~93206) obtained with the Chandra X-ray Observatory over a period
of roughly 2 years are presented. The orbit of systems A (O9.7~I+b2~v,
P$_{\rm A}=21\;$d) and B (O8~III+o9~v, P$_{\rm B}=6\;$d) are
reasonably well sampled by the observations, allowing the origin of
the X-ray emission to be examined in detail. The X-ray spectra can be
well fitted by an attenuated three temperature thermal plasma model,
characterised by cool, moderate, and hot plasma components at
$kT\simeq 0.2, 0.7,$ and 2 keV, respectively, and a circumstellar
absorption of $\simeq 0.2\times10^{22}\;$cm$^{-2}$. Although the hot
plasma component could be indicating the presence of wind-wind
collision shocks in the system, the model fluxes calculated from
spectral fits, with an average value of
$\simeq7\times10^{-13}\;$erg~s$^{-1}$~cm$^{-2}$, do not show a clear
correlation with the orbits of the two constituent binaries. A
semi-analytical model of QZ~Car reveals that a stable momentum balance
may not be established in either system A or B. Yet, despite this,
system B is expected to produce an observed X-ray flux well in excess
of the observations. If one considers the wind of the O8~III star to
be disrupted by mass transfer the model and observations are in far
better agreement, which lends support to the previous suggestion of
mass-transfer in the O8~III + o9~v binary. We conclude that the X-ray
emission from QZ~Car can be reasonably well accounted for by a
combination of contributions mainly from the single stars and the
mutual wind-wind collision between systems A and B.
\end{abstract}


\maketitle

\section{Introduction}
\label{sec:intro}
\begin{figure}
\begin{center}
\resizebox{70mm}{!}{\includegraphics{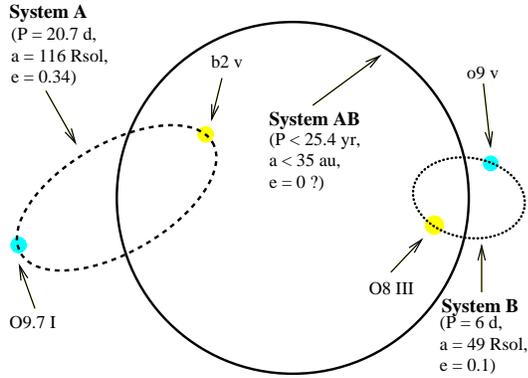}} \\
\caption[]{Schematic diagram of the multiple star system QZ~Car. For
  further details see Tables~\ref{tab:system_parameters} and
  \ref{tab:stellar_parameters}. For system AB the projected semi-major
  axis is quoted. Note that this schematic is not to scale.}
\label{fig:schematic}
\end{center}
\end{figure}

Residing within the Great Carina Nebula at a distance of 2.3 kpc
\citep[][- see also
  \citeauthor{Smith:2006}~\citeyear{Smith:2006}]{Allen:1993,
  Walborn:1995, Smith:2002}, the multiple star system QZ~Car (HD
93206) is the brightest object in the Collinder 228 star cluster, in
the older southern part of the Nebula. Via independent observations,
the presence of two systems of periodically variable lines in the
spectrum led \cite{Leung:1979} and \cite{Morrison:1979} to conclude
that there were four stars present, where the period of the stronger
line variability was due to the $\sim 20\;$day binary (hereafter
system A) and the period of the weaker lines corresponding to the
$\sim 6\;$day period eclipsing binary (hereafter system B). Despite
this great success two of the components of the system remain
undetected. The mass functions derived for the separate binary systems
suggest that the eclipsing binary component with un-detected lines is
more massive than its binary companion, whereas in the longer period,
non-eclipsing binary the unseen companion is a few times smaller than
the primary component and therefore most likely has unobservable
lines. A schematic of QZ~Car is shown in Fig.~\ref{fig:schematic} and
system and stellar parameters are noted in
Tables~\ref{tab:system_parameters} and \ref{tab:stellar_parameters},
respectively. Based on the Roche-lobe filling factors\footnote{The
  parameters in Table~\ref{tab:stellar_parameters} give Roche-lobe
  filling factors for stars B1 and B2 of $\simeq 1.0\;$and 0.4,
  respectively.}, \cite{Leung:1979} suggested that the stars in system
B have undergone some mass exchange. \cite{Morrison:1980} also noted
that there is evidence of substantial mass-loss from the primary star
in system B due to a systematic difference in velocity between He I
and Si IV, which from an evolutionary point of view makes it the most
interesting star in this system.

Little is known about the mutual orbit of systems A and B. The results
of \cite{Leung:1979} and \cite{Morrison:1980} were in agreement that
the orbital period of the super-binary must be $\ltsimm 25\;$yrs. Yet
this was based on the assumption that at the time of their
observations the system was at quadrature, or apastron in an eccentric
orbit. The speckle observations of \cite{Mason:1998} were unsuccessful
in spatially resolving the components of HD 93206, as were the more
sensitive FGS1r observations of \cite{Nelan:2004}. Therefore, the
non-resolution of the system places an upper limit of $\simeq35\;$au
on the projected separation of the two binary systems.

In this paper we report on the recent detection of X-ray emission from
QZ~Car. For single massive stars it is widely accepted that (soft)
X-ray emission is generated by embedded wind shocks (EWSs) which are
produced by the inherent instability of the line-driving mechanism
\citep[e.g.][]{Owocki:1988}. Early X-ray observations of massive stars
in binary systems revealed them to be over-luminous compared to the
expected cumulative luminosity of the separate stars
\citep[][]{Pollock:1987,Chlebowski:1991}\footnote{More recent results
  \citep{Oskinova:2005, Naze:2009, Naze:2010} suggest that only
  prominent colliding winds binary systems are significantly
  over-luminous in X-rays.}. The additional luminosity in this case is
the result of wind-wind collision shocks \citep[e.g.][]{Stevens:1992,
  Pittard:1997, Parkin:2008}. For QZ~Car, further additional X-ray
emission may be contributed by the mutual wind-wind collision between
the two binary systems (MWC). The observed flux may therefore be a
complex cocktail of X-ray emission from different sources, and
disentangling its origin(s) is not straightforward. We note that the
central multiple star system is surrounded by a subcluster of faint
X-ray emitting pre-main sequence stars, for which we refer the reader
to \cite{Townsley:2011} for a detailed analysis.
 
The X-ray observations of QZ~Car were obtained with the Chandra X-ray
observatory (hereafter \textit{Chandra}) as part of the Chandra Carina
Complex Project (CCCP)\citep{Townsley:2011, Broos:2011}. The X-ray
spectra can be reasonably well fitted by three-temperature plasma
models with a hot component at $\simeq2\;$keV. This, combined with the
fact that the observed fluxes appear to be over-luminous in comparison
to the total X-ray emission expected from the single stars, may be
indicating the presence of shock heated plasma from wind-wind
collisions. However, attempts to match the best-fit parameters from
the spectral fits to the periods of either system A or B do not reveal
any strong correlation. To aid in the interpretation of the
observations a semi-analytical model is constructed which indicates
that although normal wind-wind collision shocks are not expected,
unless the wind of the O8~III star is suppressed the model
overpredicts the observed X-ray emission by a factor of $\sim
10-20$. The suppressant in this case could be mass transfer from the
O8~III star \citep{Leung:1979}. We conclude that the dominant
contributions to the observed X-ray flux are the single stars and the
MWC. The remainder of this paper is structured as follows: in
\S~\ref{sec:observations} we present the observations,
\S~\ref{sec:results} describes the results from spectral fitting,
\S~\ref{sec:interp} details a semi-analytical model of QZ~Car. The
results from this work are discussed in \S~\ref{sec:discussion}, and
we close with a summary of our conclusions in
\S~\ref{sec:conclusions}.

\begin{table}
   \center
      \caption{System parameters for QZ Car.}
\label{tab:system_parameters}
\begin{tabular}{lllllll}
\hline
System & Components & $P$ & $a$ & $e$ & $\omega$ & $i$ \\
  & &  & (\Rsol) &  & ($^{\circ}$) & ($^{\circ}$) \\
\hline
A & A1+A2 & 20.72 d & 116 & 0.34 & 141 & 60 \\
B & B1+B2 & 5.999 d & 49 & 0.1 & $\simeq20$ & 86 \\
AB & A+B & $\ltsimm25.4\;$yrs & $\ltsimm 8687$ & 0.0 & 0 & 60 \\ 
\hline
\end{tabular}
\tablecomments{$P$ is the period of the orbit, $a$ is the semi-major
  axis of the orbit, $e$ is the orbital eccentricity, $\omega$ is the
  longitude of periastron, and $i$ is the inclination angle of the
  orbital plane (measured against the pole). Orbital periods are taken
  from \cite{Mayer:2001}, $e$'s and $\omega$'s from
  \cite{Morrison:1980}, and $i$'s from \cite{Leung:1979}. For system
  AB the projected semi-major axis is quoted. We note that $e=0.0$ is
  only a preliminary assumption for system AB, and such long-period
  systems can in fact have $0.000 \ltsimm e \ltsimm 0.999$.}
\end{table}

\section{Observations}
\label{sec:observations}

A total of nine observations over a period of roughly two years, and
combining relatively on-axis and far off-axis exposures taken with
both the I-array and the S-array, have been obtained for QZ~Car
(Table~\ref{tab:obsinfo}). Due to the brightness of the central source
the I-array observation (Obs ID 9482) was affected by photon pile-up,
which we account for in our analysis (see \S~\ref{sec:results}). The
I-array observation alone would have provided a single snapshot of
QZ~Car; however, during the CCCP QZ~Car has been observed by the
S-array CCDs on eight separate occasions, and in some cases with a
considerable exposure time (e.g. 88~ks for Obs ID 6402). Fortunately
for the current investigation this considerably expands the available
dataset.

Source and spectrum extraction were performed using \textsc{ACIS
  Extract} \citep{Broos:2002, Broos:2007, Broos:2010, Townsley:2006},
an IDL-based package developed for processing ACIS data. For each
observation, a background spectrum was taken from an annulus around QZ
Car, and the resulting background subtracted source spectrum was
binned to achieve a signal-to-noise ratio of 3. Note that the far
off-axis (ACIS-S) extractions of QZ Car encompass a nearby sub-cluster
of low-mass stars. However, these potential contaminants to the
off-axis QZ Car spectra are very weak when resolved in the on-axis
observation, and the off-axis extractions show no indication of a
large flare from one of these companions: i) the median energy shows
relatively minor changes, and, ii) Kolmogorov-Smirnov (KS) tests on
the individual lightcurves do not reveal any considerable evidence for
variability.

\section{Results}
\label{sec:results}

\begin{figure*}
  \begin{center}
    \begin{tabular}{cc}
\resizebox{60mm}{!}{\includegraphics{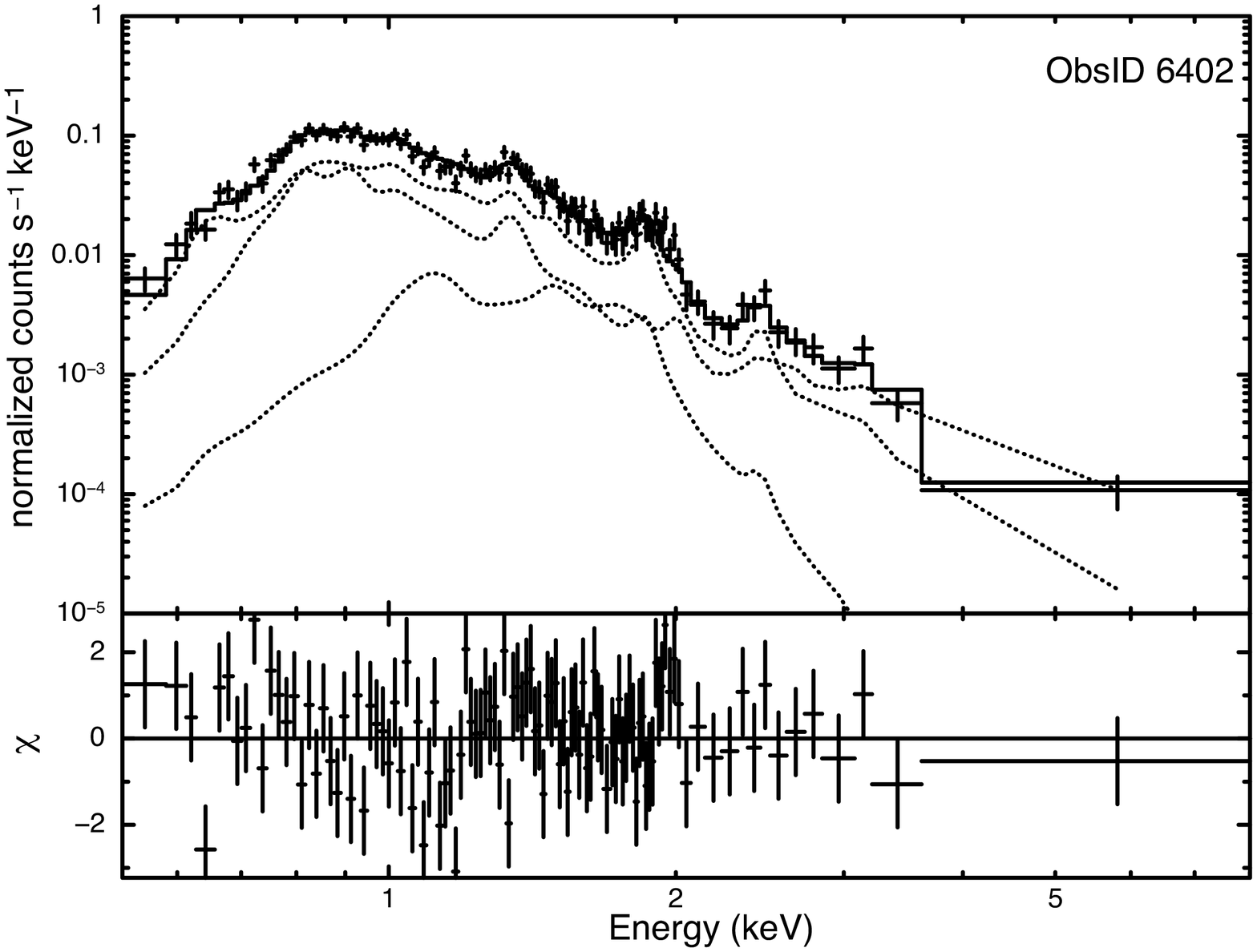}} &
\resizebox{60mm}{!}{\includegraphics{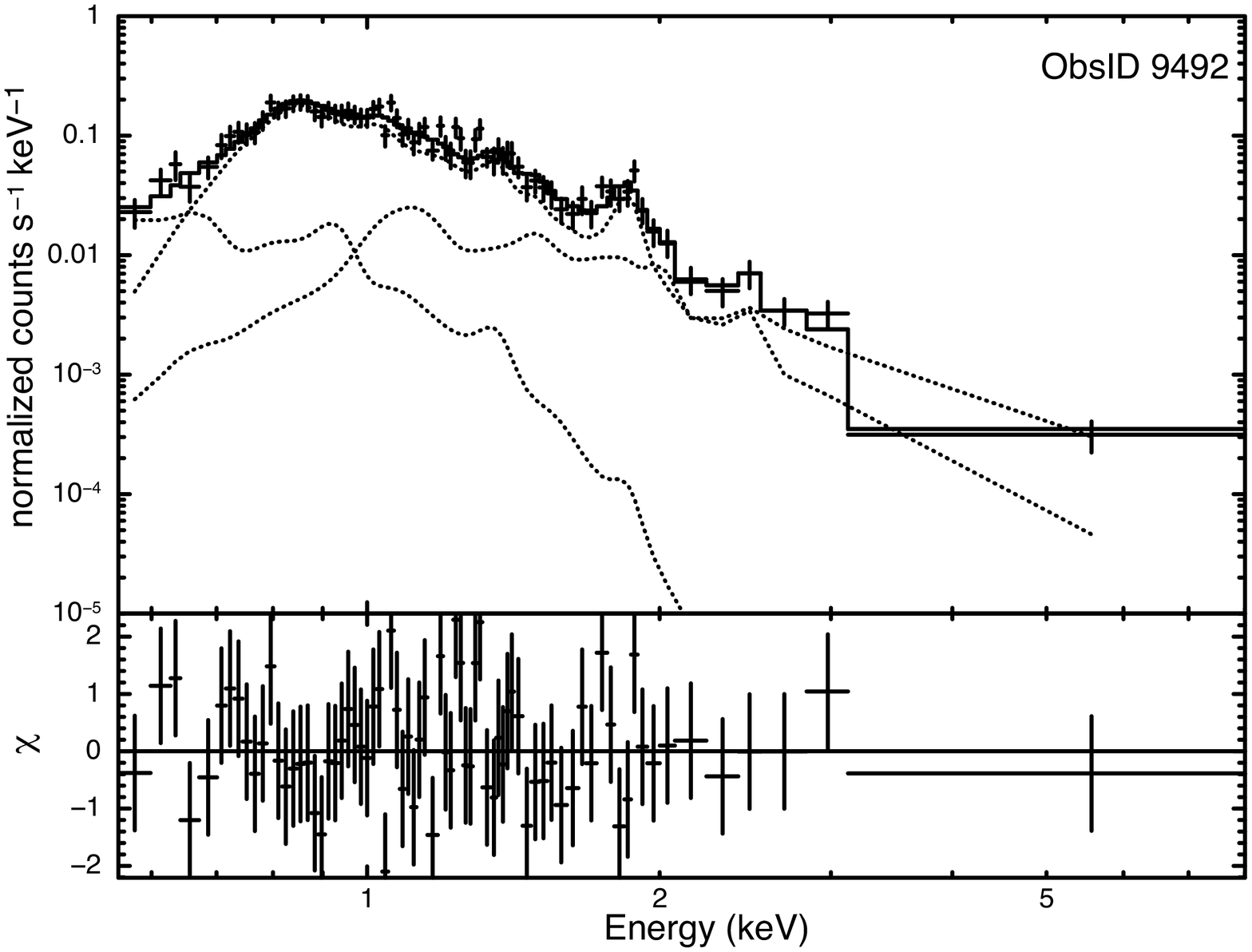}} \\

\resizebox{60mm}{!}{\includegraphics{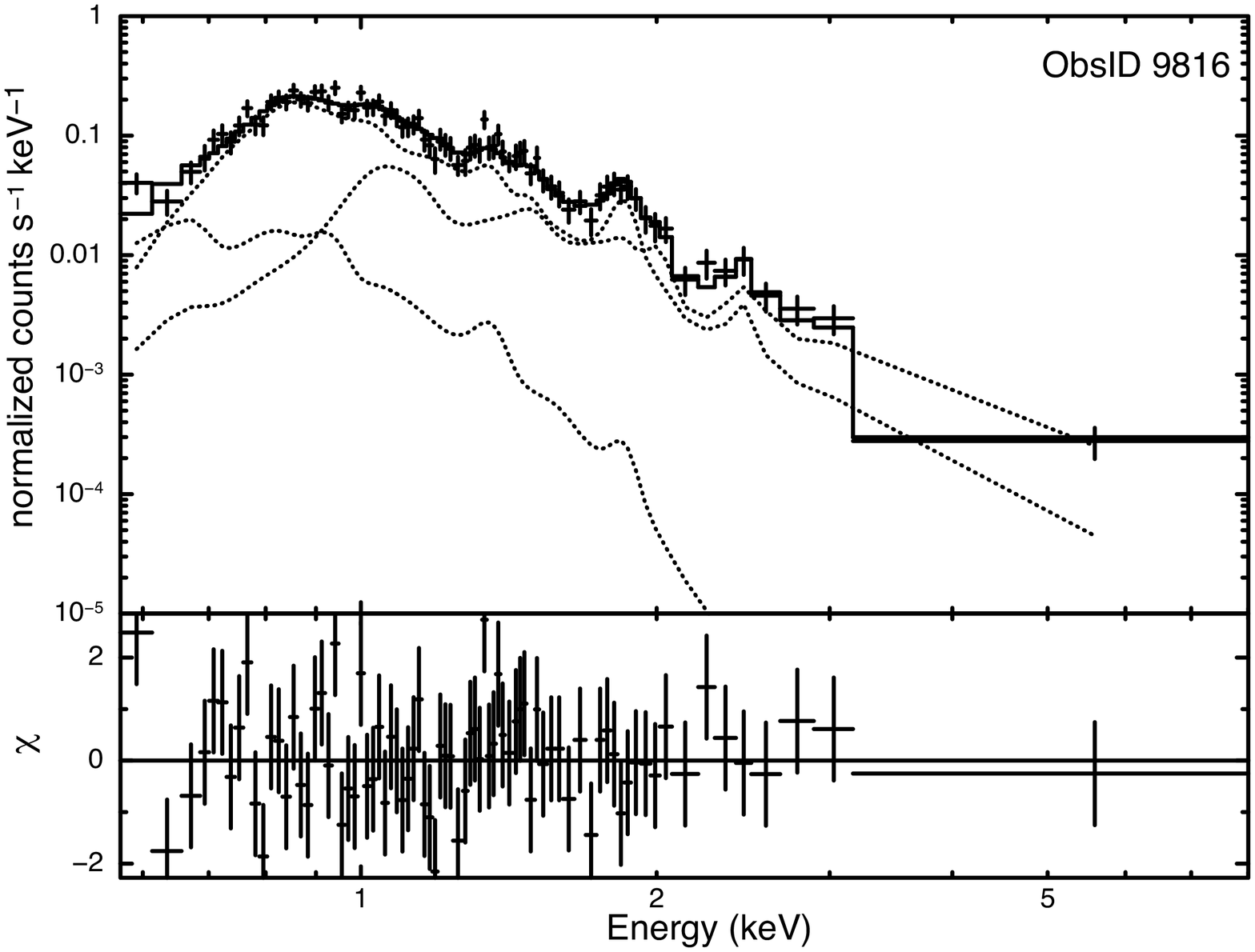}} &
\resizebox{60mm}{!}{\includegraphics{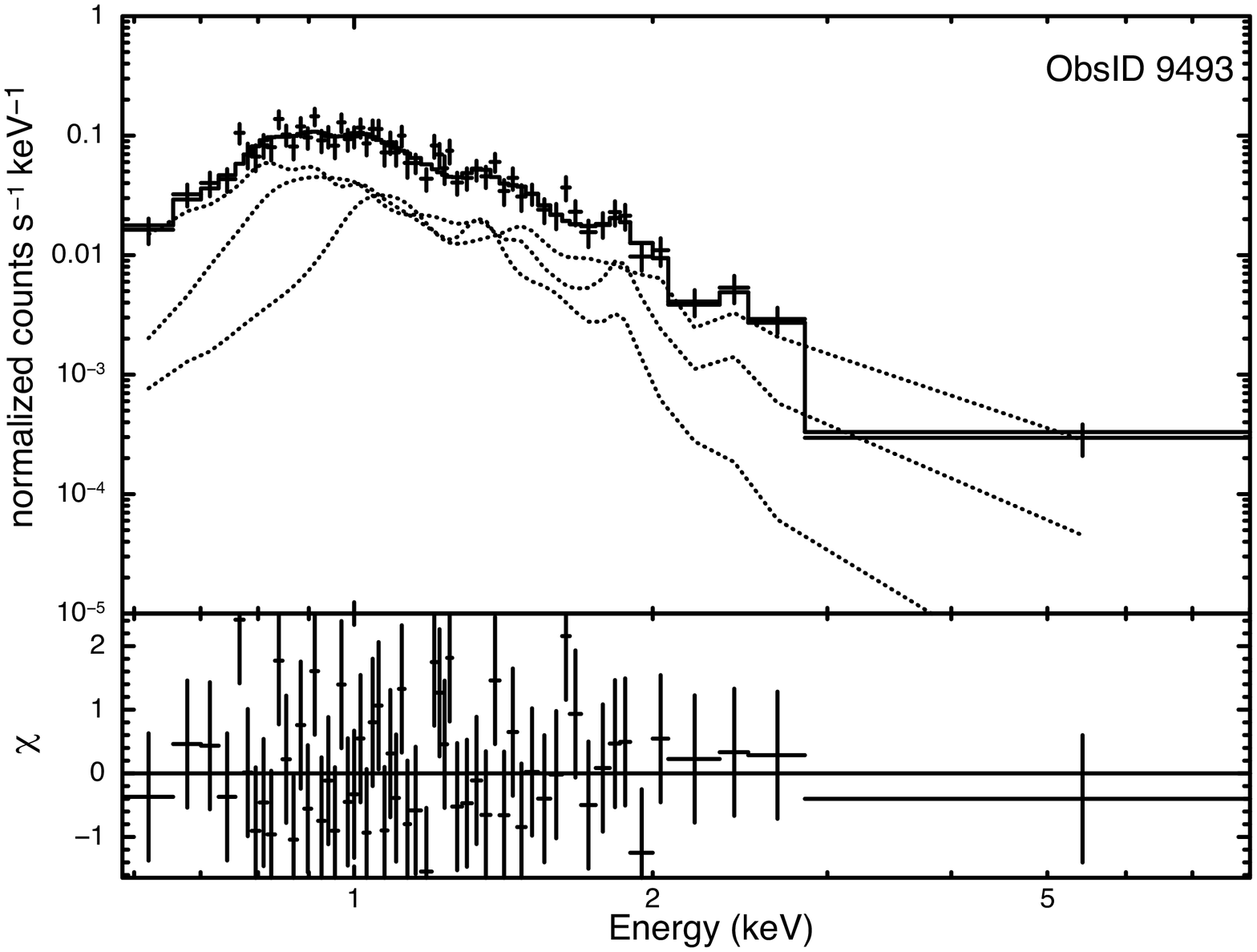}} \\

\resizebox{60mm}{!}{\includegraphics{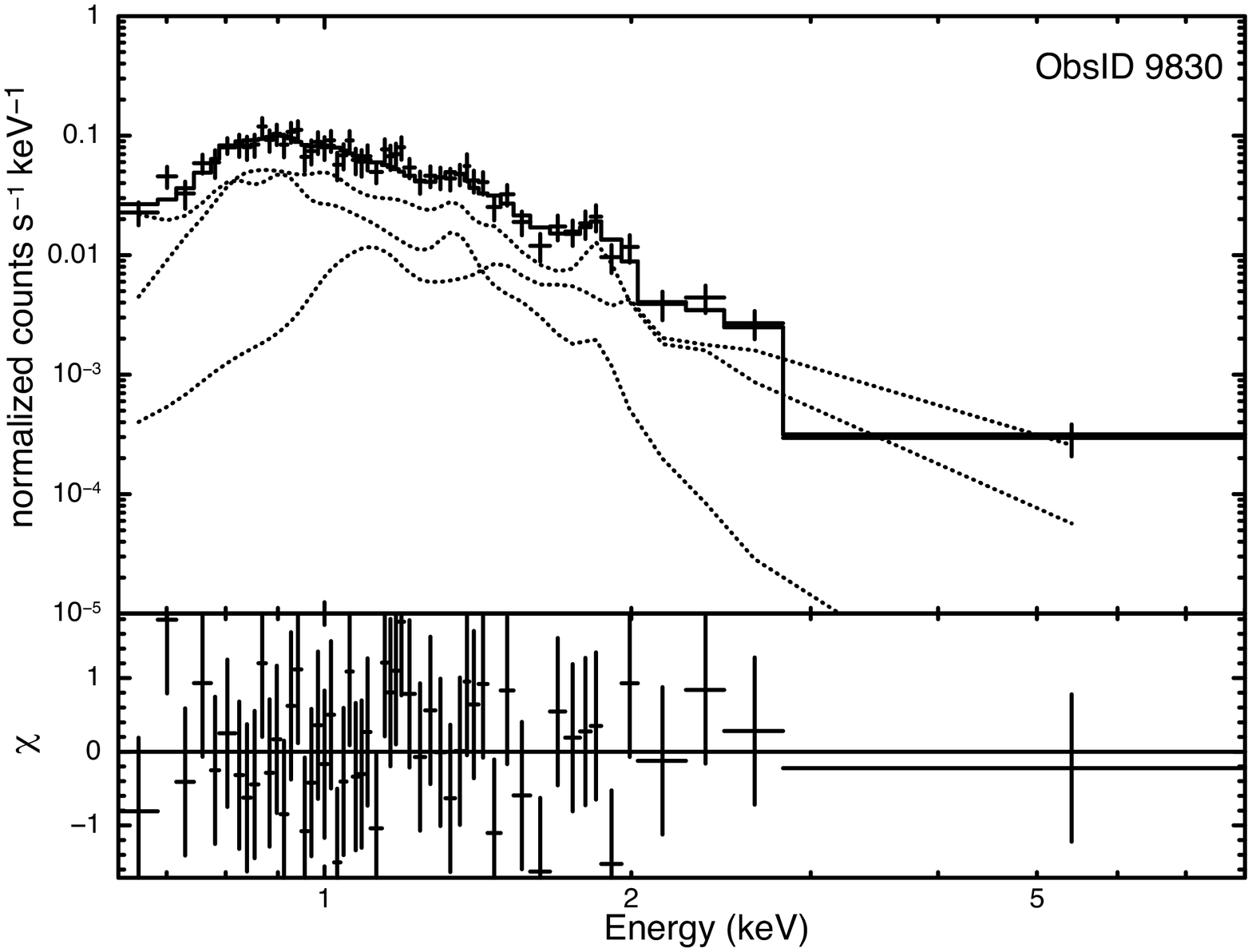}} &
\resizebox{60mm}{!}{\includegraphics{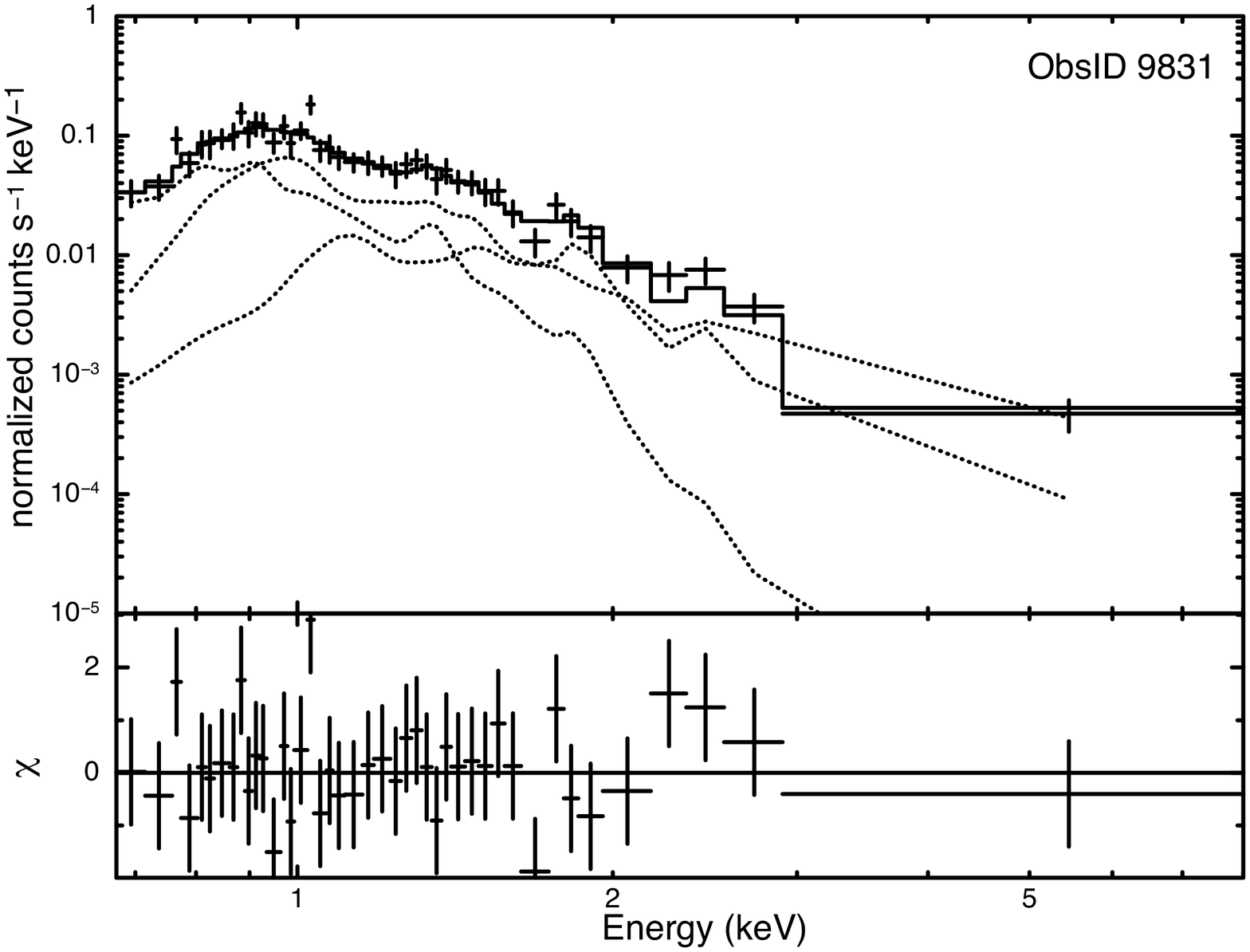}} \\

\resizebox{60mm}{!}{\includegraphics{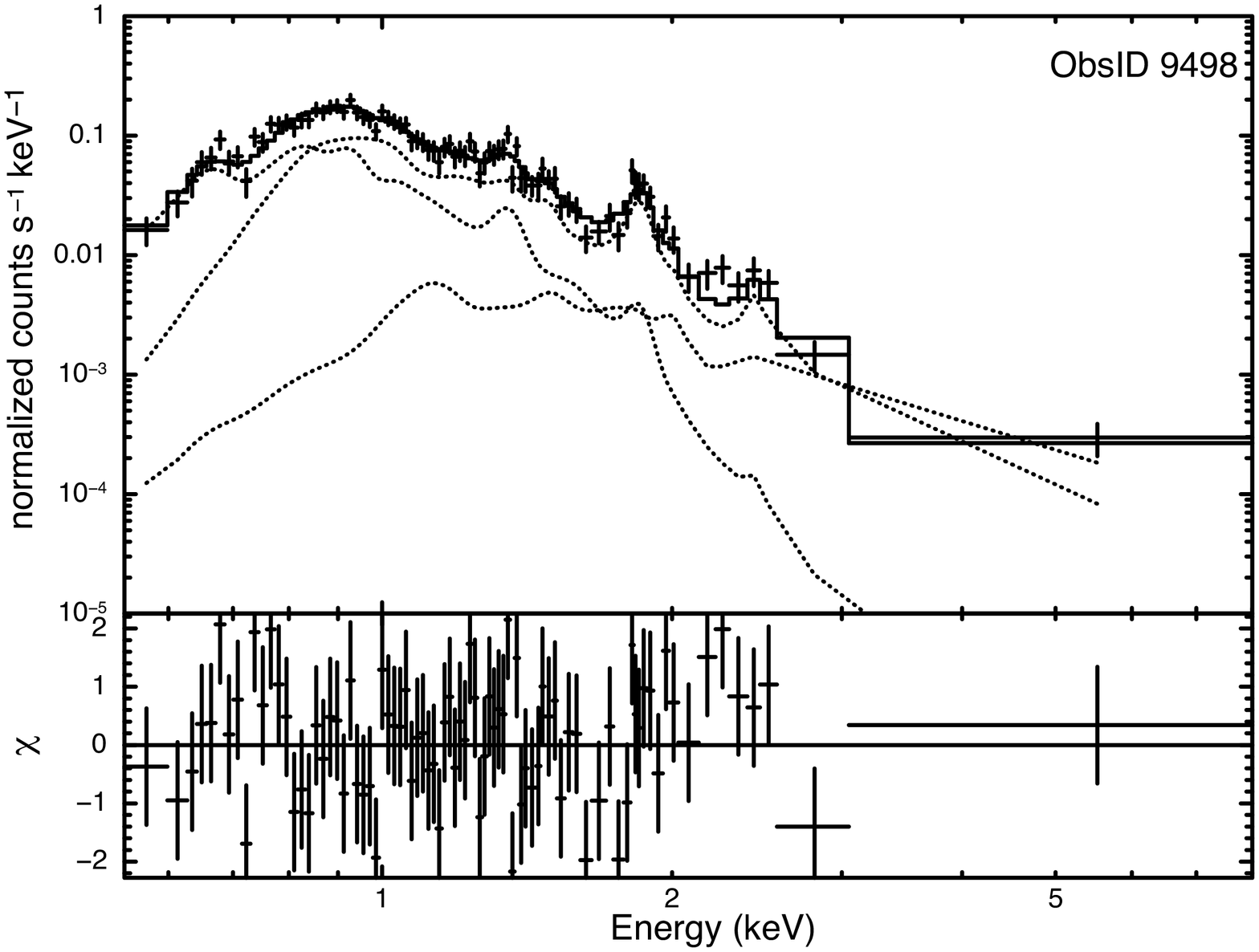}} &
\resizebox{60mm}{!}{\includegraphics{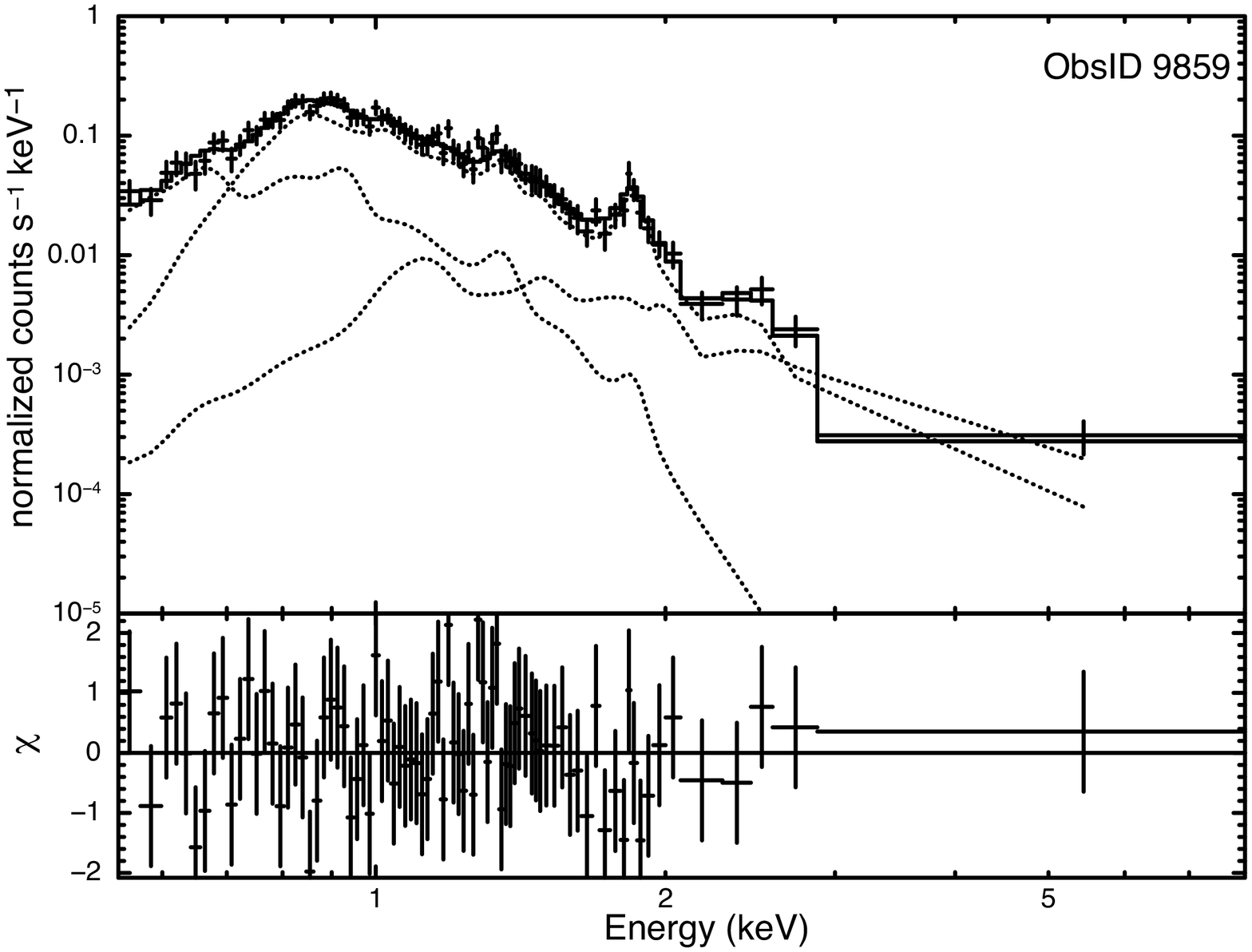}} \\

\resizebox{60mm}{!}{\includegraphics{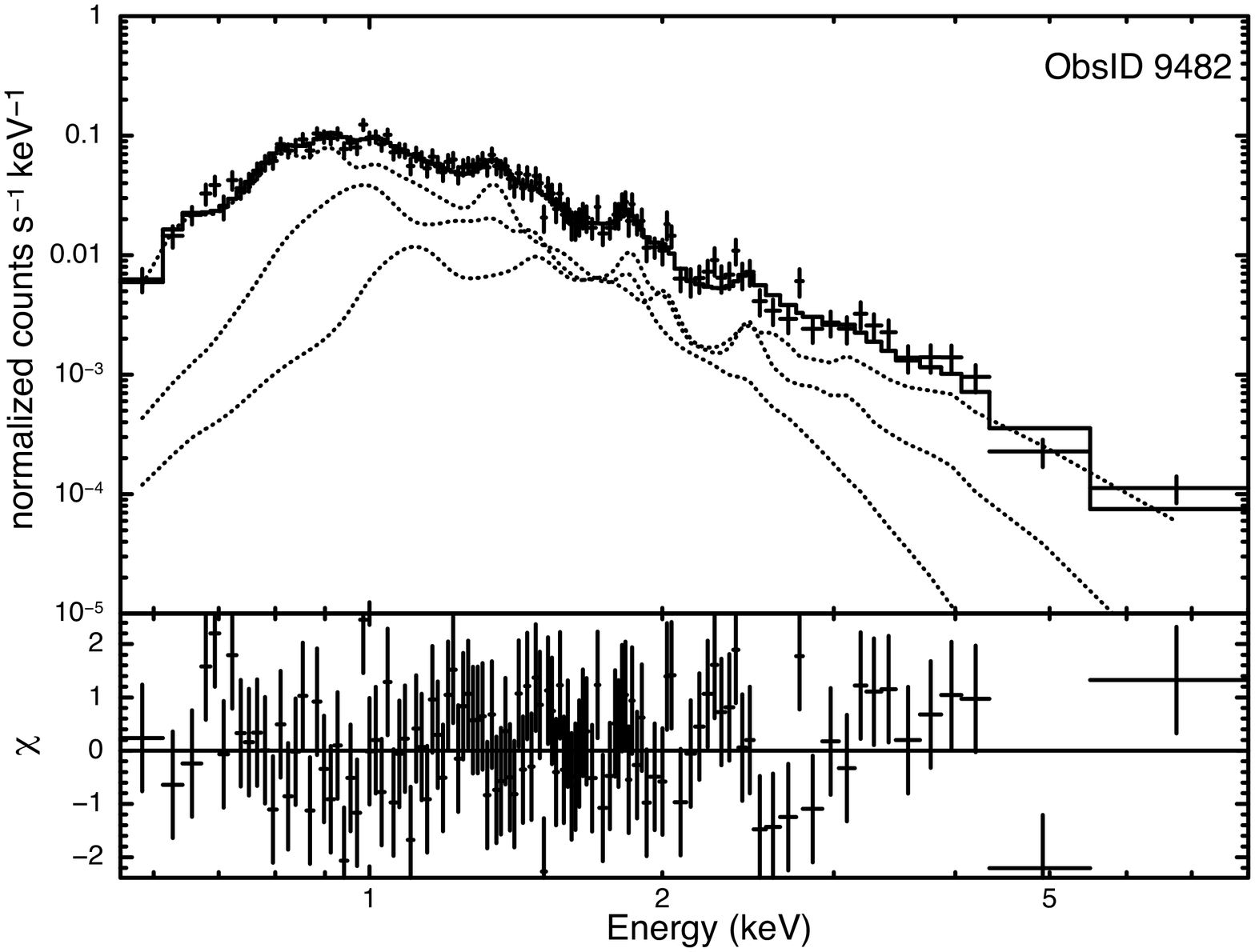}} & \\
    \end{tabular}
    \caption{0.5-8 keV X-ray spectra of QZ~Car with best fit models
      attained with the combination $tbabs_{\rm ISM}\times tbabs(apec
      + apec + apec)$.}
    \label{fig:fit5_spectra}
  \end{center}
\end{figure*}

The 0.5-8 keV spectra were fitted using v12.5.1 of {\sc
  XSPEC}\footnote{http://heasarc.gsfc.nasa.gov/docs/xanadu/xspec/}
\citep{Arnaud:1996}. To model the emission we use the {\it apec}
thermal plasma model for collisionally-ionized gas \citep{Smith:2001},
and to account for attenuation we adopted the {\it tbabs}
photoelectric absorption model \citep[see ][]{Wilms:2000}. In all
calculations the abundances were kept fixed at the solar values
\citep{Anders:1989}. To fit the spectra we use a three temperature
combination, $tbabs_{\rm ISM}\times tbabs(apec + apec + apec)$, with
an ISM absorption component, $tbabs_{\rm ISM}$, fixed at
$0.35\times10^{22}\;$cm$^{-2}$ \citep[][]{Povich:2010}. By separating
the column into ISM and circumstellar components we gain more
information about variations in the local absorption, which is a
particularly useful approach when studying the wind-wind collision in
binary systems \citep[e.g. ][]{DeBecker:2005,
  Pittard_Parkin:2010}. Similarly, the separate emission components
can be used to search for correlations with the components of the
system (i.e. EWSs, wind-wind collision regions (WCRs), and the
MWC). The physical interpretation of the adopted model combination
assumes that the circumstellar absorption to all emission components
is the same\footnote{A more physically meaningful model combination
  would incorporate separate absorption components for each emission
  component (i.e. spatially distinct emission regions). However, due
  to limited statistics attempts to fit the spectra with such a model
  combination were unsuccessful.}. As noted in
\S~\ref{sec:observations}, Obs ID 9482 was affected by
\anchorfoot{http://cxc.harvard.edu/ciao/why/pileup\_intro.html}{photon
  pile-up}, which we correct for using an additional pileup model
\citep{Davis:2001}.  The pileup parameter {\em fr\_time} was frozen to
3.31 s to account for exposure time discarded during the standard data
cleaning process.\footnote{ The
  \anchorparen{http://cxc.harvard.edu/ciao/download/doc/pileup\_abc.pdf}{Chandra
    ABC Guide to Pileup} describes the correction to {\em fr\_time}
  that is required for all ACIS sources.  } ; the best-fit values for
the thawed $\alpha$ and {\em psffrac} parameters were found to be
0.954 and 0.633.

A visual inspection of the spectra shows that a three temperature
thermal plasma model provides a reasonably good fit to the data
(Fig.~\ref{fig:fit5_spectra}). In general there is a cool component at
$kT_1\simeq 0.2\;$keV, a moderate temperature component at
$kT_2\simeq0.7\;$keV, a hot component at $kT_3\simeq 2\;$keV, a column
density, $N_{\rm H}\simeq 0.2\times10^{22}\;$cm$^{-2}$, and a 0.5-8
keV flux $\simeq 7\times10^{-13}\;$erg~s$^{-1}$~cm$^{-2}$. Comparing
these results to those for QZ Car in \cite{Naze:2010} we see that
there is agreement in the circumstellar column and in the presence of
a moderate temperature component at $\simeq0.7\;$keV. However, the
derived flux and temperature of the hotter plasma component are
notably higher in this work. These differences are likely due to the
use of three-temperature fits in this work, whereas
\citeauthor{Naze:2010} apply a two-temperature fit\footnote{The
  quality of the data constrains the complexity of the model that can
  be applied in spectral fitting. \cite{Naze:2010} adopted
  two-temperature model fits to perform a consistent analysis of the
  entire OB star sample from the CCCP, which consists of data of
  varying quality. In the present paper, we aim to perform a more
  detailed analysis of QZ Car, for which the data is of sufficient
  quality to permit meaningful results from the use of a
  three-temperature model.}. We note that two-temperature fits were
also examined, however, statistically better results could be attained
for all observations using three-temperature spectral fits. For
example, for ObsID 6402, the reduced chi-squared attained from
two-temperature and three-temperature fits were 1.84 and 1.33,
respectively. In both cases the highest temperature component was at
$kT \simeq 2.1\;$keV. The higher chi-squared in the case of the
two-temperature model was due to poorer fit to the spectrum at
energies $\ltsimm 1\;$keV. Adding the third temperature component
significantly remedied this.

From the best-fit parameters in Table~\ref{tab:fit5_values} (see also
Fig.~\ref{fig:fit5_plots}) one sees that $kT_1$ and $kT_2$ appear to
be reasonably well constrained. However, in constrast $kT_3$ and
$N_{\rm H}$ do not. Recalling that $N_{\rm H}$ is intended to account
for the circumstellar absorption to potentially numerous regions of
X-ray emitting plasma, this is unsurprising. Despite this, using the
conversion factor of $N_{\rm H}=A_{\rm v}\times
(1.9\times10^{21})\;$cm$^{-2}$ \citep[e.g. ][]{Cox:2000}, there is
good agreement with the visual extinction from the observed colour
indices \citep[$A_{\rm v}=1.2\;$mag; ][]{Herbst:1976,
  Leung:1979}. Comparisons of the X-ray spectra of QZ~Car against
two-temperature fits to single and binary stars of similar spectral
type in Carina \citep{Naze:2010} are inconclusive in so much as they
do not directly support/rule-out the presence of wind-wind collision
shocks based solely on the spectral shape and the derived plasma
temperatures (i.e. the presence of a hot component with $kT\gtsimm
1\;$keV).

Examining the variation of the best-fit parameters plotted against the
orbital phases of the binary systems we see that both $kT_1$ and
$kT_2$ remain relatively constant across all observations
(Fig.~\ref{fig:fit5_plots}). A tentative correlation between $kT_3$
and system A could be suggested although the errors are quite large
(due to limited statistics in the spectra at higher energies) and one
could equally favour a null result. Additionally, an increase in
$N_{\rm H}$ would be expected as higher temperature plasma close to
the apex of the WCR(s) comes into view, which is not seen in the
fits. Finally, we note that the 0.5-8 keV fluxes calculated from the
spectral fits do not show any clear correlation with the orbit of
either system A or B.

\begin{table*}
   \center
      \caption{Stellar parameters for QZ Car.}
\label{tab:stellar_parameters}
\begin{tabular}{llllllll}
\hline 
Component & Sp. Type & $T_{\rm eff}$ & $R_{\ast}$ & $M_{\ast}$ & $L_{\ast}$ & $\dot{M}$ & $v_{\infty}$  \\
          &  & (K)           & (\Rsol) & (\Msol) & (log$[L_{\ast}/\Lsol]$) & ($\Msolpyr$) & (km s$^{-1}$) \\ 
\hline
A1 & O9.7~I & 32000 & 22.5 & 40 & 5.7 & $2.2\times10^{-6}$ & 2140 \\
A2 & b2~v & 20000 & 6.0 & 10 & 3.7 & $2.4\times10^{-9}$ & 1040 \\ 
B1 & 08~III & 32573 & 26.9 & 14.1 & 5.3 & $5.2\times10^{-7}$ & 2220 \\
B2 & o9~v & 32463 & 8.9 & 28 & 4.9 & $6.4\times10^{-8}$ & 2850 \\
\hline
\end{tabular}
\tablecomments{For consistency we adopt the labelling of
  \cite{Mayer:2001} for the system components. The spectral types for
  components A1 and B1 were determined from observations by the OWN
  Team (R.~Barba, private communication). The values of $T_{\rm eff}$,
  $R_{\ast}$, and $M_{\ast}$ for components A1 and B2 are taken from
  \cite{Leung:1979}, for component B1 values were taken from
  \cite{Martins:2005}, and for component A2 the respective values have
  been estimated by a comparison to objects of similar spectral type
  in \cite{Prinja:1989}. The values of $v_{\infty}$ are calculated as
  $2.6~v_{\rm esc}$ for components A1, B1, and B2, and as $1.3~v_{\rm
    esc}$ for component A2 (based on $T_{\rm eff}$), where $v_{\rm
    esc}=\sqrt{2GM_{\ast}/R_{\ast}}$. The $\dot{M}$ values are
  calculated using the cooking recipe from \cite{Vink:2000}. We note
  that the spectral types of components A2 and B2 are not based on
  true spectral classifications (i.e. direct detection in the optical
  spectrum), and are in fact based on photometric and/or colour
  information. Therefore, we use lowercase letters to denote the
  spectral type of components A2 and B2.}
\end{table*}

\begin{table}
 \center
  \caption{Summary of the observations.}
\label{tab:obsinfo}
\begin{tabular}{llllllll}
\hline
Obs ID	& CCD & $\theta$ & Date & $T$ & $R$ & $\phi_{\rm A}$ & $\phi_{\rm B}$ \\
 &  & ($'$) &  &  (ks) & (ks$^{-1}$) &  &  \\
\hline
6402 & S2 & 16.01 & 30 Aug 2006 & 87 & 76.5 & 0.10 & 0.07 \\
9492 & S3 & 14.04 & 12 Feb 2008 & 20 & 119.0 & 0.68 & 0.50 \\
9816 & S3 & 14.04 & 15 Feb 2008 & 21 & 138.5 & 0.81 & 0.94 \\
9493 & S2 & 17.23 & 25 Feb 2008 & 20 & 83.1 & 0.32 & 0.70 \\
9830 & S2 & 17.23 & 28 Feb 2008 & 20 & 73.3 & 0.45 & 0.14 \\
9831 & S2 & 17.23 & 1 Mar 2008 & 16 & 78.4 & 0.52 & 0.41 \\
9498 & S3 & 18.94 & 24 May 2008 & 32 & 118.3 & 0.61 & 0.53 \\
9859 & S3 & 18.94 & 31 May 2008 & 28 & 129.4 & 0.91 & 0.57 \\
9482 & I3 & 3.41 & 18 Aug 2008 & 57 & 73.3 & 0.75 & 0.83 \\
\hline
\end{tabular}
\tablecomments{$\phi_{\rm A}$ and $\phi_{\rm B}$ are the corresponding
  orbital phases for system A and system B, respectively, calculated
  using the ephimerides of \cite{Mayer:2001}. The time of periastron
  used for system A is taken as JD 2442530.49, which refines the time
  of periastron determined by \cite{Morrison:1980} using the more
  recent observations of \cite{Mayer:2001}. For system B, the
  ephemeris is extrapolated from the minimum observed by
  \cite{Mayer:1998} at JD 2448687.16. $\theta$ is the off-axis angle,
  $T$ is the exposure time, and $R$ is the count rate for each
  observation. The CCCP ACIS source label and official source name for
  QZ~Car are C2\_1111 and 104422.91-595935.9, respectively.}
\end{table}

\section{A model of QZ~Car}
\label{sec:interp}

The results of the analysis so far do not highlight any obvious link
between the observed X-ray emission and the orbit of either
binary. However, with potentially multiple sources of X-rays it is
possible that any signature of the orbit may be smeared-out in the
cumulative emission. To better constrain the r\^{o}les of the various
components of the system, and their contribution to the total
emission, we now construct a semi-analytical model of QZ~Car.

\begin{table*}
 \center
  \caption{Results from spectral fitting.}
\label{tab:fit5_values}
\begin{tabular}{llllllllll}
\hline Obs ID & $N_{\rm H 1}$ & $kT_{1}$ &  Norm$_1$ & $kT_{2}$ & Norm$_2$ & $kT_{3}$ & Norm$_3$ &  0.5 - 8 keV flux & $\chi^{2}\;$ \\
 & ($10^{22}\;$cm$^{-2}$) & (keV) & ($10^{-4}\;$cm$^{-5}$) & (keV) & ($10^{-4}\;$cm$^{-5}$) & (keV) & ($10^{-4}\;$cm$^{-5}$) &  ($10^{-13}\;$erg cm$^{-2}$ s$^{-1}$) &  (d.o.f) \\
\hline 
6402 & 0.25$^{0.05}_{0.05}$ & 0.26$^{0.05}_{0.06}$ & 23.4$^{12.9}_{8.4}$ & 0.60$^{0.12}_{0.04}$ & 7.22$^{3.06}_{3.82}$ & 2.10$^{1.12}_{0.46}$ & 1.32$^{0.45}_{0.49}$ & 6.55 & 1.33 (102) \vspace{1mm}\\
9492 & 0.20$^{0.09}_{0.10}$ & 0.15$^{0.05}_{0.06}$ & 24.3$^{94.8}_{18.3}$ & 0.58$^{0.03}_{0.03}$ & 9.71$^{2.42}_{2.80}$ & 1.90$^{1.12}_{0.40}$ & 2.52$^{0.78}_{0.93}$ & 7.09 & 0.96 (68) \vspace{1mm}\\
9816 & 0.12$^{0.08}_{0.08}$ & 0.20$^{\rm fr}$ & 5.46$^{7.08}_{3.74}$ & 0.61$^{0.04}_{0.03}$ & 7.73$^{2.24}_{1.16}$ & 1.49$^{0.20}_{0.23}$ & 3.32$^{0.62}_{0.72}$ & 7.70 & 1.00 (74)\vspace{1mm}\\
9493 & 0.16$^{0.14}_{0.15}$ & 0.28$^{0.07}_{0.08}$ & 15.0$^{27.0}_{11.1}$ & 0.69$^{0.31}_{0.12}$ & 3.37$^{5.33}_{2.11}$ & 1.49$^{0.47}_{0.26}$ & 3.20$^{1.05}_{1.21}$ & 7.00 & 0.97 (52) \vspace{1mm}\\
9830 & 0.24$^{0.15}_{0.17}$ & 0.23$^{0.08}_{0.08}$ & 26.8$^{88.8}_{20.2}$ & 0.62$^{0.15}_{0.10}$ & 5.81$^{5.05}_{4.10}$ & 1.94$^{2.92}_{0.48}$ & 1.96$^{1.02}_{0.99}$ & 6.19 & 0.76 (47) \vspace{1mm}\\
9831 & 0.19$^{0.21}_{0.19}$ & 0.23$^{0.13}_{0.23}$ & 27.0$^{160}_{24.2}$ & 0.79$^{0.25}_{0.17}$ & 4.57$^{4.20}_{2.22}$ & 2.37$^{4.65}_{0.80}$ & 2.80$^{1.22}_{1.40}$ & 7.84 & 0.94 (36) \vspace{1mm}\\
9498 & 0.27$^{0.09}_{0.08}$ & 0.23$^{0.03}_{0.05}$ & 34.9$^{32.3}_{15.9}$ & 0.71$^{0.05}_{0.10}$ & 7.16$^{4.28}_{1.85}$ & 2.63$^{-}_{1.3}$ & 1.05$^{0.87}_{0.74}$ & 7.14 & 1.14 (78) \vspace{1mm}\\
9859 & 0.22$^{0.10}_{0.09}$ & 0.18$^{0.06}_{0.04}$ & 42.3$^{45.3}_{22.1}$ & 0.58$^{0.03}_{0.03}$ & 10.8$^{2.97}_{4.36}$ & 2.14$^{-}_{0.76}$ & 1.32$^{0.76}_{0.67}$ & 7.78 & 0.77 (76)\vspace{1mm}\\
9482 & 0.31$^{0.06}_{0.08}$ & 0.26$^{0.03}_{0.03}$ & 31.3$^{26.8}_{16.1}$ & 0.81$^{0.13}_{0.11}$ & 2.66$^{2.80}_{0.85}$ & 1.96$^{0.35}_{0.31}$ & 1.63$^{0.11}_{0.81}$ & 5.28 & 1.03 (106)\vspace{1mm}\\
\hline \vspace{0.5mm}
Average & 0.22 & 0.22 & 25.6 & 0.67 & 6.56 & 2.00 & 2.12 & 6.95 & 0.99 \vspace{0.5mm}\\
\hline
\end{tabular}
\tablecomments{The ``fr'' indicates that the parameter was frozen at
  this value during spectral fitting. The $90\%$ confidence level
  errors are quoted with a hyphen corresponding to an unconstrained
  error.}
\end{table*}

\subsection{Intrinsic X-ray emission}
\label{subsec:int_emission}

Firstly, we estimate the contribution to the X-ray luminosity from
embedded wind shocks (EWSs) using the canonical relation $L_{\rm X} =
10^{-7} L_{\rm bol}$, which amounts to a flux of $\simeq
4.4\times10^{-13}\;$erg~s$^{-1}$~cm$^{-2}$. Then, following
\cite{Pittard:2002}, we can estimate the X-ray flux from the wind-wind
collisions using the simple relation:
\begin{eqnarray}
f_{\rm X i} = \frac{1}{8 \pi D^2} \dot{M_{\rm i}}v_{\rm i}^{2} \frac{\Xi_{\rm i}}{\chi_{\rm i}} \label{eqn:lx}
\end{eqnarray}

\noindent where $D$ is the distance to QZ~Car (taken to be 2.3 kpc),
$\Xi$ is the fractional wind kinetic power normal to the contact
discontinuity, $\chi$ is the cooling parameter \citep[=
  $v_{8}^{4}d_{12}/\dot{M}_{-7}$, see][]{Stevens:1992}, and the
subscript $i$ denotes the contributing component. The parameter $\Xi$
is dependent on the wind momentum ratio of the system, $\eta_{\rm
  ij}=(\dot{M}_{\rm j} v_{\rm j})/(\dot{M}_{\rm i} v_{\rm i})$, where
the subscript $j$ denotes the component index of the binary
companion. For $\eta_{\rm ij}$ = (0.01, 0.1, 1.0), $\Xi_{\rm i} =
(0.0042, 0.033, 0.167)$ and $\Xi_{\rm j} = (0.564, 0.403, 0.167)$,
i.e. the value of $\Xi$ is higher for the weaker wind because a
greater fraction of that wind collides close to the shock normal. The
parameter $\chi$ is the ratio of the characteristic flow time to the
cooling time; if $\chi \ltsimm 1$ the post-shock gas is radiative,
whereas if $\chi \gg 1$ the post-shock gas is adiabatic. Note that in
the case where $\chi < 1$ we set $\chi = 1$ to satisfy energy
conservation.

In systems A and B the separation of the stars is relatively small,
and the stellar winds may not have reached their terminal
velocities. This has consequences for the position of the momentum
balance surface between the stars (if one exists). If we assume that
the wind velocity follows a $\beta$-velocity law (i.e. $v(r) =
v_{\infty}(1- R_{\ast}/r)^{\beta}$), and set $\beta= 1$ for
simplicity, we can calculate the position of the momentum balance
point by numerical solution of the following equation for the distance
from star i to the momentum balance point (along the line-of-centres),
$r$,
\begin{eqnarray}
  \frac{\dot{M}_{\rm i} v_{\rm \infty i}}{r^{2}}\left ( 1 - \frac{R_{\rm \ast i}}{r}\right ) = 
  \frac{\dot{M}_{\rm j} v_{\rm \infty j}}{(d_{\rm sep} - r)^{2}}\left (1 - \frac{R_{\rm \ast j}}{d_{\rm sep} - r}\right ).
\end{eqnarray}

\noindent It is then straight-forward to calculate the effective
values of $\eta$, $\Xi$, $\chi$, and $f_{\rm X i}$ from the shocked
gas of each wind, where the preshock wind speed along the
line-of-centres rather than the terminal wind speed is used in the
calculations. The distance from the star to the shock rather than the
binary separation is used to calculate $\chi$.

We can repeat this process for the MWC. For this purpose we
approximate the mass-loss rates and terminal wind speeds for systems A
and B as,
\begin{eqnarray}
  \dot{M}_{\rm A} & = & \dot{M}_{\rm A1} + \dot{M}_{\rm A2},\label{eqn:mdot_A}\\ 
  \dot{M}_{\rm B} & = & \dot{M}_{\rm B1} + \dot{M}_{\rm B2},\label{eqn:mdot_B}\\
  v_{\rm A} & = & \left(\frac{\dot{M}_{\rm A1}v_{\rm A1}^{2}  + \dot{M}_{\rm A2}v_{\rm A2}^{2}}
  {\dot{M}_{\rm A1} + \dot{M}_{\rm A2}}  \right)^{1/2}, \label{eqn:vel_A}\\
  v_{\rm B} & = & \left(\frac{\dot{M}_{\rm B1}v_{\rm B1}^{2}  + \dot{M}_{\rm B2}v_{\rm B2}^{2}}
  {\dot{M}_{\rm B1} + \dot{M}_{\rm B2}}  \right)^{1/2}. \label{eqn:vel_B}
\end{eqnarray}

\noindent Parameter values pertaining to the X-ray emission
calculations for the MWC are listed in Table~\ref{tab:mutual_wwc}. We
note that Eqs~\ref{eqn:mdot_A}-\ref{eqn:vel_B} should provide a
reasonable approximation as, due to the relatively large separation of
system AB, the stellar winds should have had sufficient time to mix.

At all orbital phases in systems A and B a wind-wind momentum balance
does not occur and the wind of the weaker star is completely crushed
by the stronger opponent\footnote{The weaker star may radiatively
  brake the incoming wind, permitting a ram pressure balance
  \citep{Gayley:1997}. This is unlikely to be effective for system A
  (due to the relatively low luminosity of component A2 compared to
  component A1), but may be effective in system B.}. Therefore, the
values of $\Xi_{\rm A2}$ and $\Xi_{\rm B2}$ are set to zero (i.e. no
contribution to the X-ray emission) and $\Xi_{\rm A1}$ and $\Xi_{\rm
  B1}$ are instead approximated as the fractional solid angle
subtended by the face of their respective companion star,
\begin{eqnarray}
  \Xi_i = \frac{1}{4\pi}
  \left(1 - \cos \left\{ \tan^{-1}\left(\frac{R_{\rm \ast j}}{d_{\rm sep i-j}}\right)\right\}\right). \label{eqn:solid_angle}
\end{eqnarray}

Fig.~\ref{fig:wind-wind_plots} shows the variation of the pre-shock
velocities, $\chi$'s, $\Xi$'s, and $kT$ with orbital phase. Note that,
to avoid confusion with references to the intrinsic emission from the
individual stars, the characteristics of component A1's wind colliding
against its opposing star (component A2) is referred to as component
A1-O. The same nomenclature is adopted for component B1. The
characteristic energy of the emitted X-rays is given by $kT \simeq
1.17 v_{8}^{2}\;$keV, where $v_{8}$ is the pre-shock velocity in units
of $10^{8}\;$cm s$^{-1}$. For now, the preshock velocities of
component B1-O are calculated assuming that it drives a wind towards its
binary companion, rather than the system being semi-detached. Later
(\S~\ref{subsec:comparison2obs}) we consider the possibility of zero
colliding winds emission from system B.

Evidently, terminal wind speeds are not reached prior to collision
(see Table~\ref{tab:stellar_parameters}). The cooling parameters,
$\chi$'s, for components A1-O and B1-O are sufficiently high for the
post-shock gas to be adiabatic at phases close to apastron, whereas
they may become radiative ($\chi \ltsimm 1$) around periastron. The
value of $\Xi$ for component B1-O is the highest at all orbital
phases, representative of the larger fractional solid angle subtended
by its companion star in comparison to component A1-O (see
Eq~\ref{eqn:solid_angle}). The separation of systems A and B is
sufficiently large that for the MWC the stellar winds will have
reached their terminal velocities when they collide and this factor,
combined with the low post-shock densities, leads to adiabatic shocks
($\chi_{\rm A} = 570$ and $\chi_{\rm B} = 2860$).

\begin{table}
   \center
      \caption{Parameters for the mutual wind-wind collision between
        systems A and B.}
\label{tab:mutual_wwc}
\begin{tabular}{lll}
\hline
Parameter & System A & System B \\
  & &  \\
\hline
$\dot{M}$ (\Msolpyr) &  $2.2\times10^{-6}$ & $5.8\times10^{-7}$ \\
$v_{\infty}$ (cm s$^{-1}$) & $2.14\times10^{8}$  & $2.30\times10^{8}$ \\
$\Xi$ & 0.06 & 0.35 \\
$\chi$ & 570 & 2860 \\
$f_{\rm X}$  ($10^{-13}\;$erg~s$^{-1}$~cm$^{-2}$) & 5.3  & 1.9 \\
$\eta$ &  \multicolumn{2}{c}{0.28} \\ 
$d_{\rm sep (A+B)}$ ($10^{12}$ cm) & \multicolumn{2}{c}{600}  \\
$N_{\rm H}$ (cm$^{-2}$)& \multicolumn{2}{c}{$1.3\times10^{20}$} \\
\hline
\end{tabular}
\tablecomments{The values of $f_{\rm X}$ and $N_{\rm H}$ are
  calculated from Eqs.~\ref{eqn:lx} and \ref{eqn:nh},
  respectively. The values of $\Xi$ for System A and System B were
  interpolated from the results of \cite{Pittard:2002} for the
  respective values of $\eta$.}
\end{table}

\begin{figure}
 \begin{center}
   \begin{tabular}{c}
     \resizebox{85mm}{!}{\includegraphics{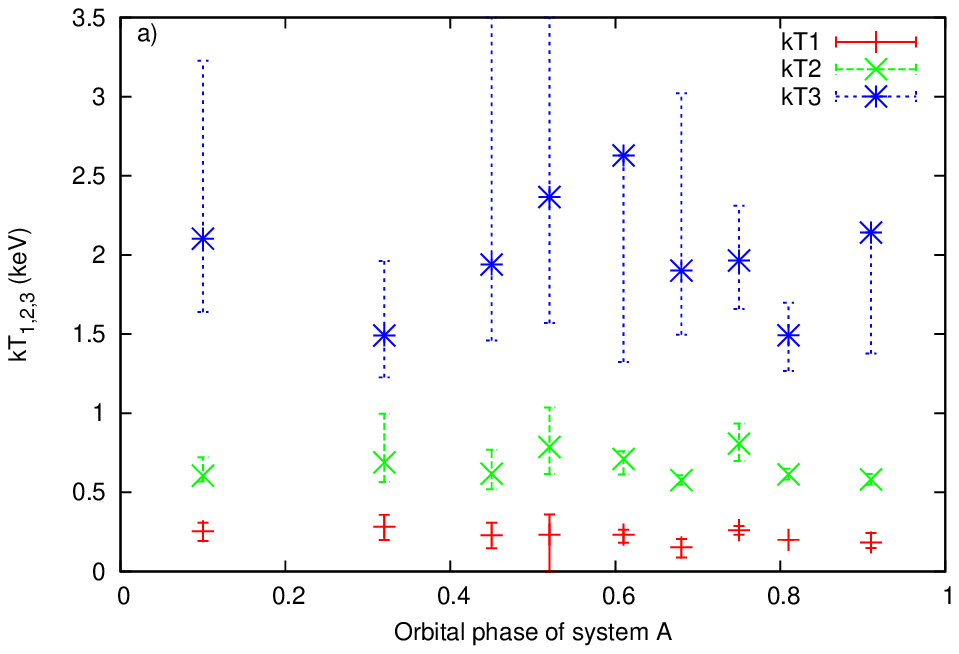}} \vspace{-4mm} \\
     \resizebox{85mm}{!}{\includegraphics{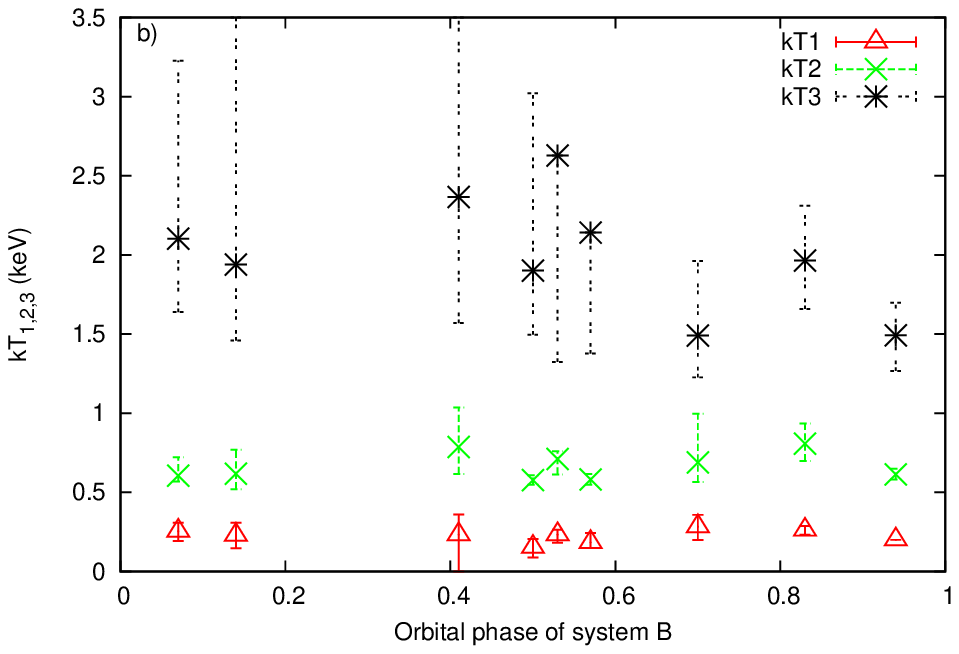}} \vspace{-4mm} \\
      \resizebox{85mm}{!}{\includegraphics{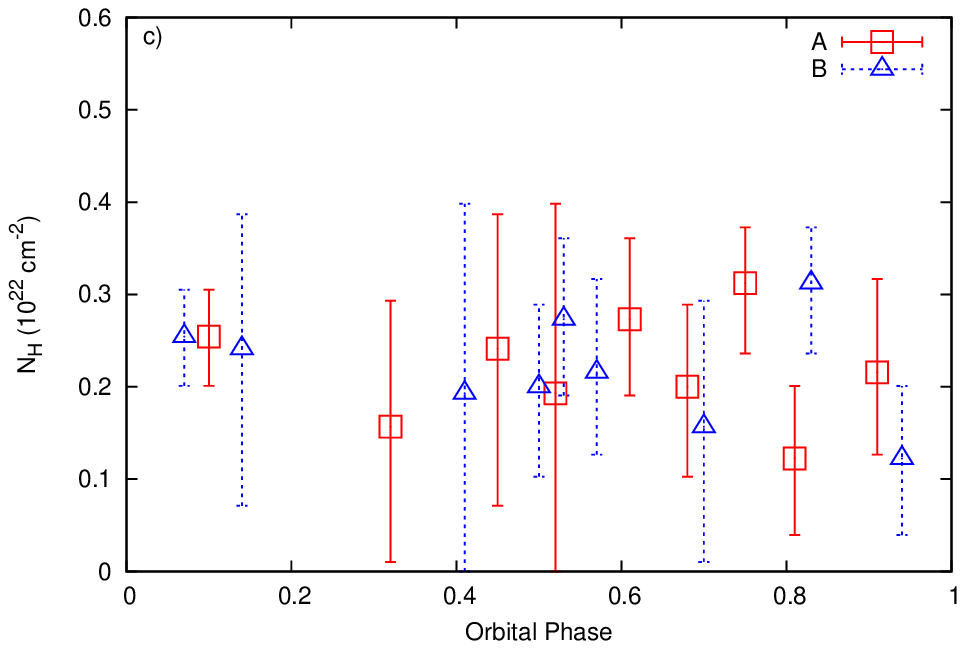}} \vspace{-4mm}\\
      \resizebox{85mm}{!}{\includegraphics{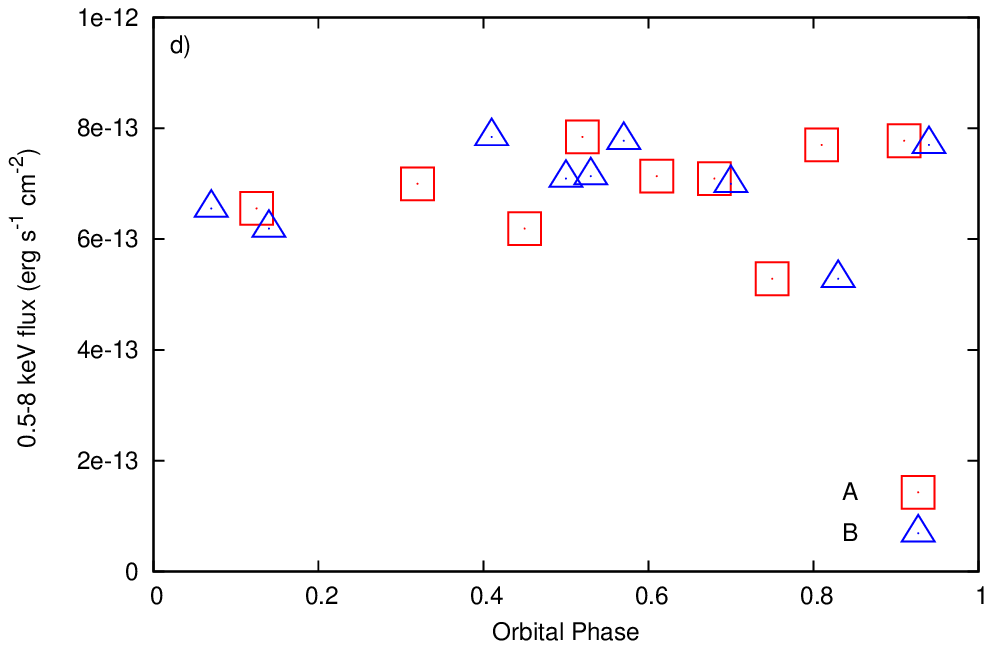}} \vspace{-1mm} \\
    \end{tabular}
    \caption{Optimal parameters attained from the spectral fits
      plotted against orbital phase for both system A and system
      B. From top to bottom: a) $kT$'s for system A, b) $kT$'s for
      system B, c) $N_{\rm H}$, and d) the 0.5-8 keV flux. See also
      Table~\ref{tab:fit5_values}.}
    \label{fig:fit5_plots}
  \end{center}
\end{figure}

With a range of pre-shock velocities, the X-ray spectrum for QZ Car
may well be dominated by emission from different shocked plasma
components at different energies. Approximating the mean $kT$ to be
roughly half of the maximum value (to account for shock obliquity
downstream from the apex of the WCR) we see that the predicted values
for component A1-O (Fig.~\ref{fig:wind-wind_plots}d) are slightly
lower than those derived for the hot plasma component from the
spectral fits (Fig.~\ref{fig:fit5_plots} and
Table~\ref{tab:fit5_values}) which has a mean temperature of
2.00~keV. In contrast, the mean temperatures from the MWC, where
$kT$'s are $\simeq 2.7$ and 3.1~keV for system A and B respectively,
are higher than the observationally determined value. We note that all
the values are higher than our adopted mean plasma temperature for
EWSs of $kT=0.25\;$keV \citep[e.g. ][]{Owocki:1999}.

The intrinsic X-ray flux from the individual shocked winds is shown in
Fig.~\ref{fig:wind-wind_plots}e. Component B1-O has the highest
intrinsic X-ray luminosity, and so system B will dominate the X-ray
emission if one assumes that the WCR has not been disrupted by mass
transfer (for an alternative scenario see
\S~\ref{subsec:comparison2obs}). Component A1-O is the {\em faintest}
emitter (noting that component A1-O is bright for a brief period around
periastron which is not sampled by our observations - see the
$\phi_{\rm A}$'s in Table~\ref{tab:obsinfo}), followed by the EWSs
($\approx 4.4 \times10^{-13}\;$erg~s$^{-1}$~cm$^{-2}$), both of which
are fainter than the contribution from the MWC
(Table~\ref{tab:mutual_wwc}). However, before making a detailed
comparison with the best-fit values from the spectral fits we can
improve our predictions by considering the energy dependence of the
intrinsic and attenuated flux.

\subsection{Attenuated emission}
\label{subsec:att_emission}

\begin{figure}
 \begin{center}
   \begin{tabular}{c}
     \resizebox{70mm}{!}{\includegraphics{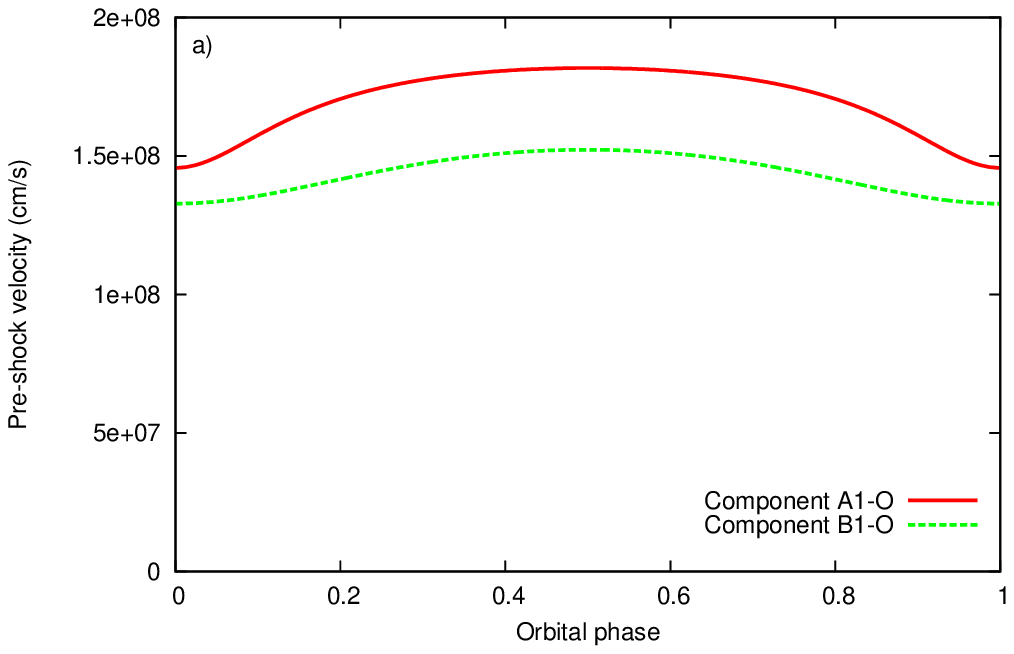}} \vspace{-5.5mm} \\
\resizebox{70mm}{!}{\includegraphics{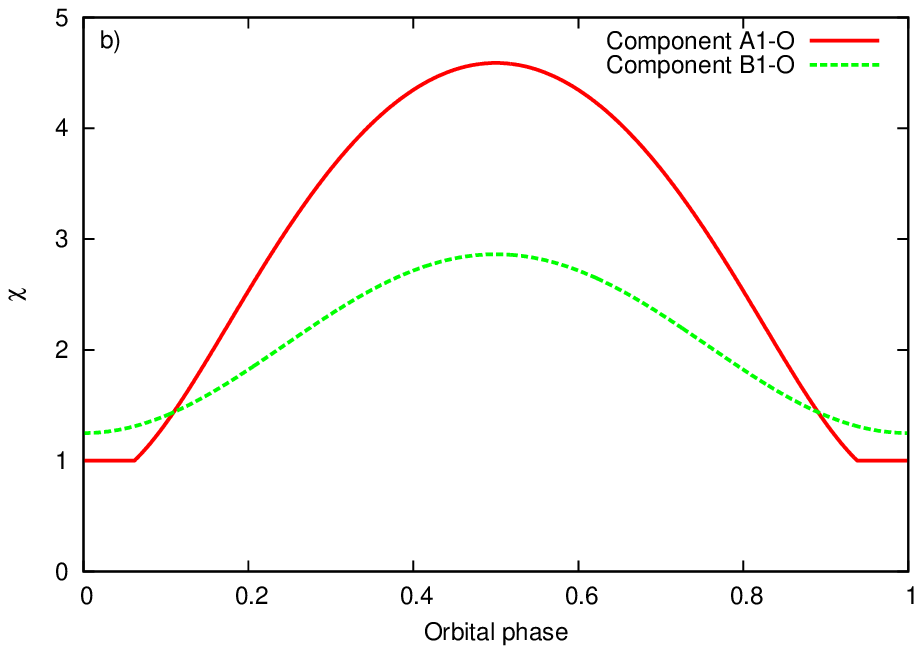}} \vspace{-5.5mm} \\
      \resizebox{70mm}{!}{\includegraphics{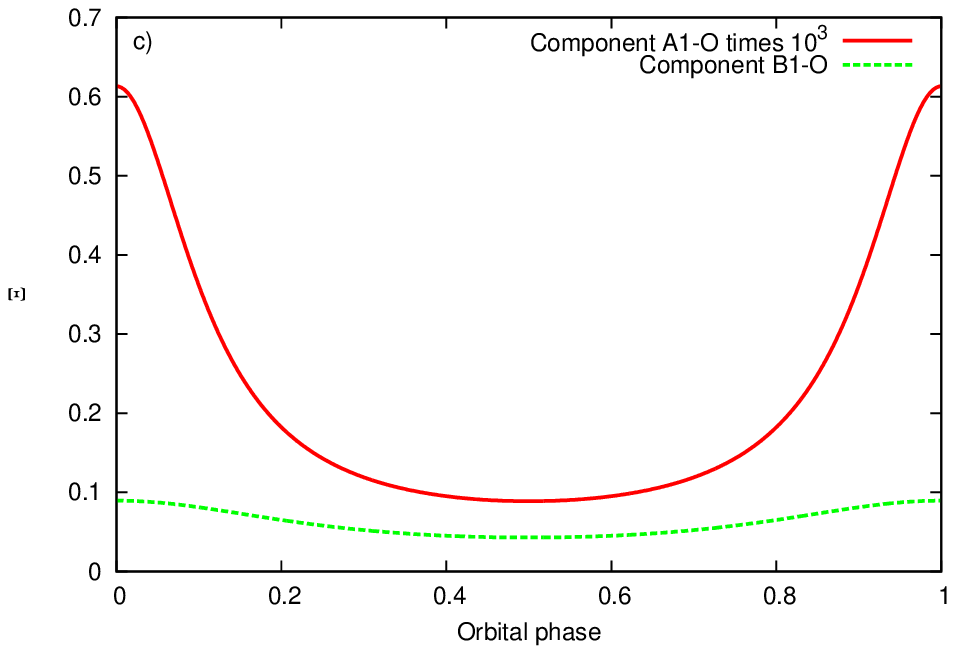}} \vspace{-5.5mm}\\
\resizebox{70mm}{!}{\includegraphics{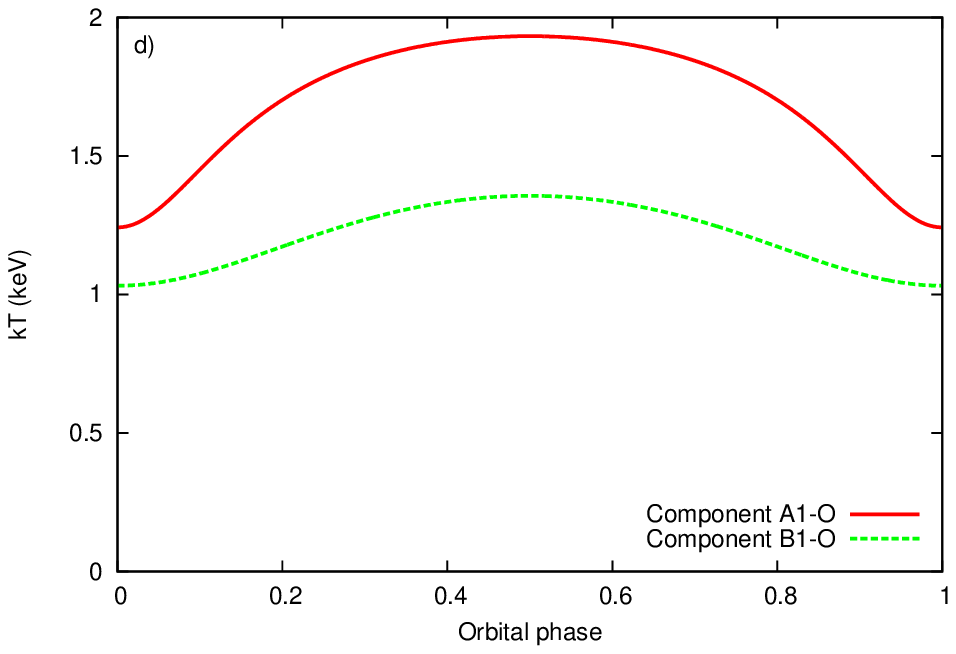}} \vspace{-5.5mm}\\
     \resizebox{70mm}{!}{\includegraphics{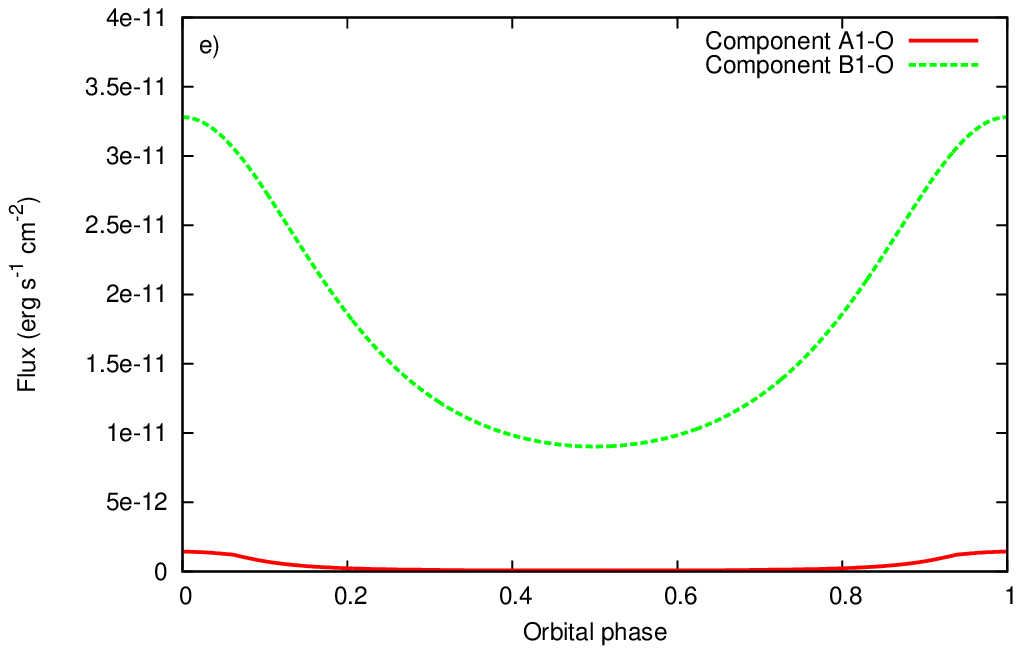}} \vspace{-2mm}\\
    \end{tabular}
    \caption{Variation with orbital phase of various parameters in the
      calculations of the X-ray luminosity generated in the multiple
      wind-wind collisions. From top left to bottom right: a)
      pre-shock velocity, b) cooling parameter $\chi$, c) fraction of
      wind kinetic power thermalized in the collision $\Xi$, d) the
      {\it mean} post-shock temperature $kT$, and e) the intrinsic
      X-ray flux due to wind-wind collisions from each
      component. $\chi$'s are limited to 1 if the actual value is
      lower. The contributions from the post-shock gas of the
      different components are phased to their respective system,
      i.e. the contribution from component A1-O is phased to system
      A. Note that in the plot $\Xi_{\rm A1}$ has been multiplied by a
      factor of $10^{3}$ for comparative purposes.}
    \label{fig:wind-wind_plots}
  \end{center}
\end{figure}

\begin{figure}
 \begin{center}
   \begin{tabular}{c}
     \resizebox{80mm}{!}{\includegraphics{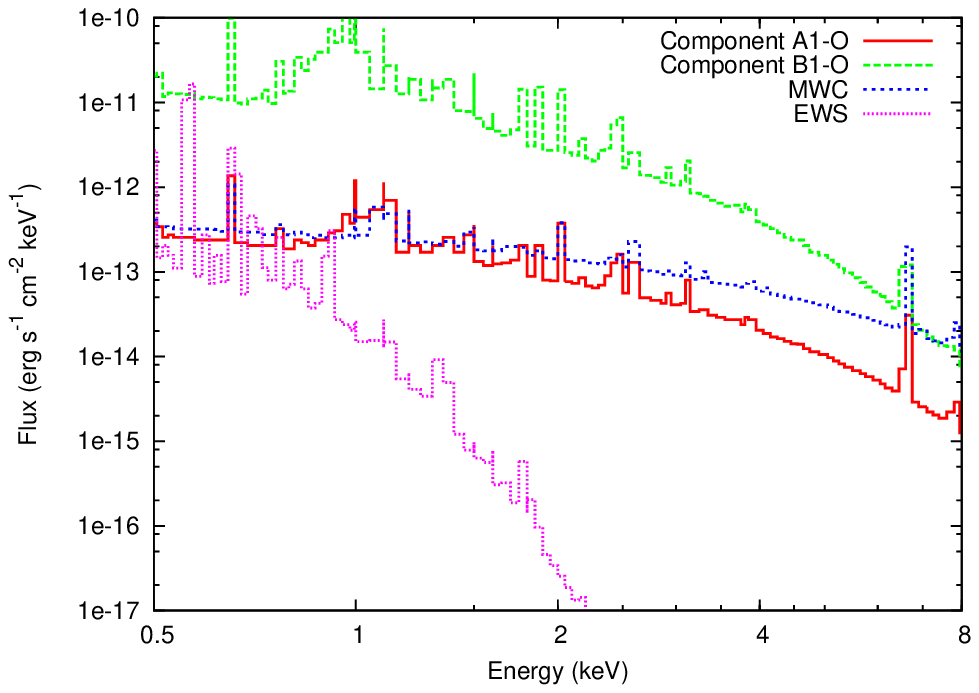}} \\
     \resizebox{80mm}{!}{\includegraphics{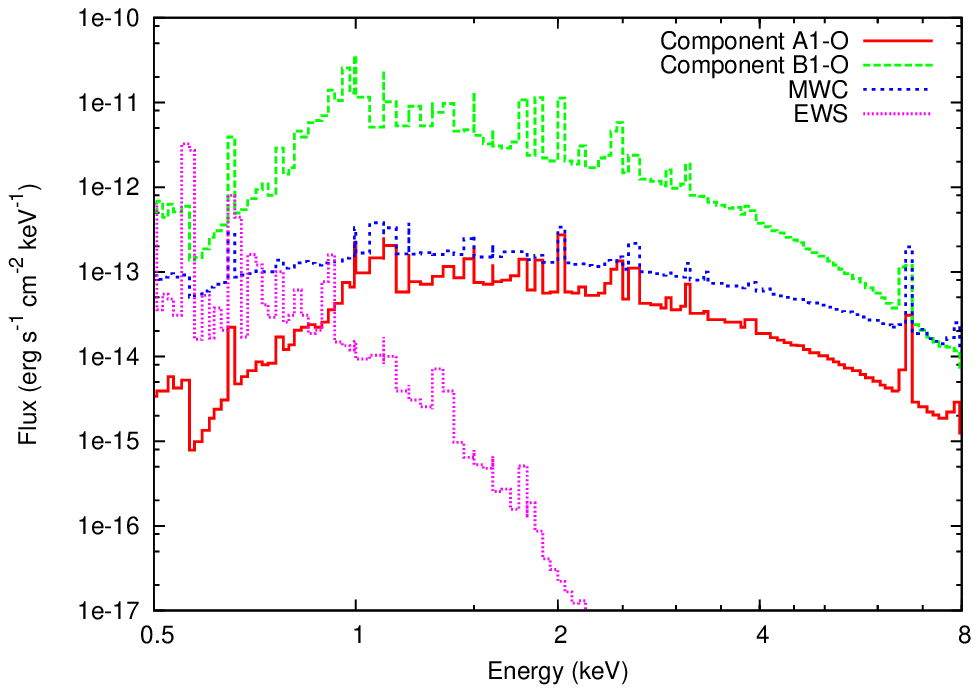}} \\
    \end{tabular}
    \caption{Intrinsic (top panel) and attenuated (bottom panel)
      synthetic 0.5-8~keV X-ray spectra for Obs ID 6402.}
    \label{fig:lx_specs}
  \end{center}
\end{figure}

The range of preshock velocities evident in
Fig.~\ref{fig:wind-wind_plots}a will cause the spectra from the
different components to have different characteristic energies. We can
examine the implications of this energy dependence by firstly
calculating an intrinsic spectrum at the mean post-shock gas
temperature using the MEKAL plasma code \citep{Kaastra:1992,
  Mewe:1995}, where solar abundances are assumed
\citep{Anders:1989}. Each intrinsic spectrum is then scaled so that
the integrated 0.5-8~keV fluxes are equal to the orbital phase
dependent values, which for components A1-O and B1-O are shown in
Fig.~\ref{fig:wind-wind_plots}e. The top panel of
Fig.~\ref{fig:lx_specs} shows the intrinsic spectra calculated for
components A1-O, B1-O, the EWSs, and the MWC for Obs ID 6402 (see
Table~\ref{tab:obsinfo} for the respective orbital phases of systems A
and B). The EWSs clearly contribute the softest spectrum. For the
wind-wind collision shocks, the lower preshock velocity for component
B1-O relative to the MWC and component A1-O also results in a slightly
softer spectrum. Interestingly, although not the brightest emitter,
the MWC has the hardest spectrum.

To estimate the impact of circumstellar absorption, a characteristic
column density for the binary systems can be calculated using Eq(11)
from \cite{Stevens:1992},

\begin{equation}
  \bar{N}_{\rm H} = 5\times10^{21}\frac{\dot{M}_{-7}}{v_{8}}
  \frac{(1 + \eta^{1/2})}{d_{12}} \label{eqn:nh}
\end{equation}

\noindent The expression for $\bar{N}_{\rm H}$ is for a binary system
at quadrature where the winds are assumed to be at their terminal
velocities ($v_{\infty}$'s are used to calculate $\eta$ in this
case). When the stars are at quadrature all lines of sight to the
emitting region will pass through the more powerful wind, which is
assumed to be the dominant absorber. The orbital phase dependent
characteristic column densities for systems A and B are $\simeq
4.9-9.9 \times10^{21}\;$cm$^{-2}$ and $\simeq 4.5-5.6
\times10^{21}\;$cm$^{-2}$, respectively. Note that in using
Eq~\ref{eqn:nh} we are essentially assuming the system is always at
quadrature. To calculate the optical depth, we then multiply the total
column density (=$\bar{N}_{\rm H} + N_{\rm H~ISM}$) by the opacity for
gas at $10^{4}\;$K calculated using version $c08.00$ of Cloudy
\citep[][see also \cite{Ferland:1998}]{Ferland:2000}, where solar
abundances are assumed \citep{Anders:1989}. It is important to
highlight that no circumstellar absorption is added for the EWSs as
the $L_{\rm X}/L_{\rm bol}$ relation is for sources which have been
corrected only for ISM absorption. Coincidently, the circumstellar
absorption to the WCR between systems A and B is very small, so that
this component essentially suffers only ISM absorption. The resulting
attenuated spectra are displayed in the bottom panel of
Fig.~\ref{fig:lx_specs}. With the exception of the EWS emission, there
is a consistent trend of a turnover energy of $\sim
1\;$keV. Furthermore, the biggest victims of energy dependent
absorption are components A1-O and B1-O, which can be seen from a
comparison of the intrinsic and attenuated spectra in
Fig.~\ref{fig:lx_specs}.

\begin{figure}
 \begin{center}
   \begin{tabular}{c}
     \resizebox{80mm}{!}{\includegraphics{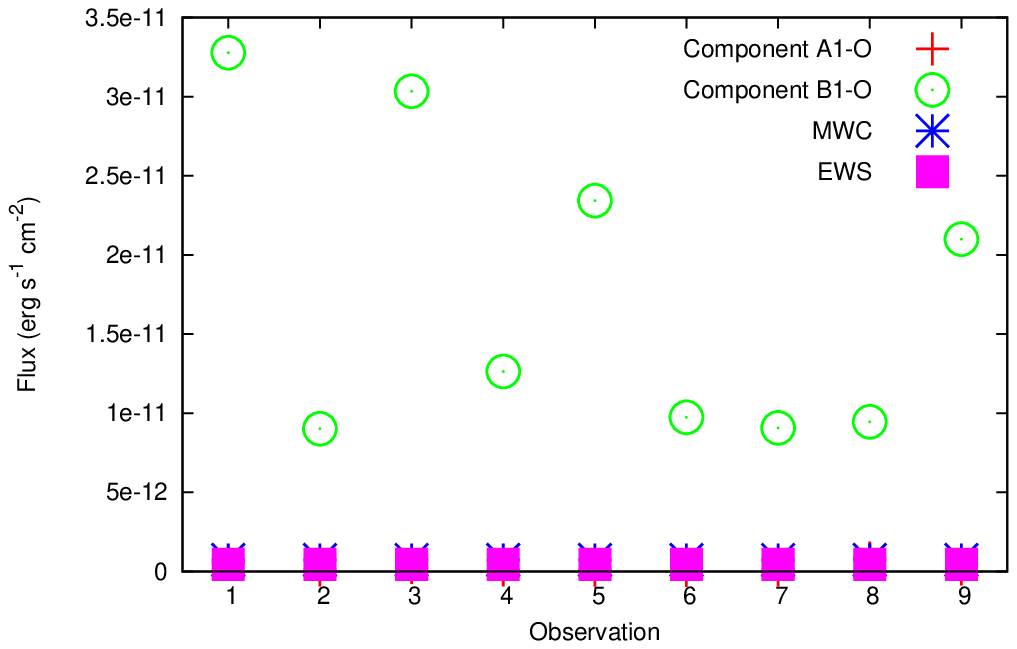}} \\
     \resizebox{80mm}{!}{\includegraphics{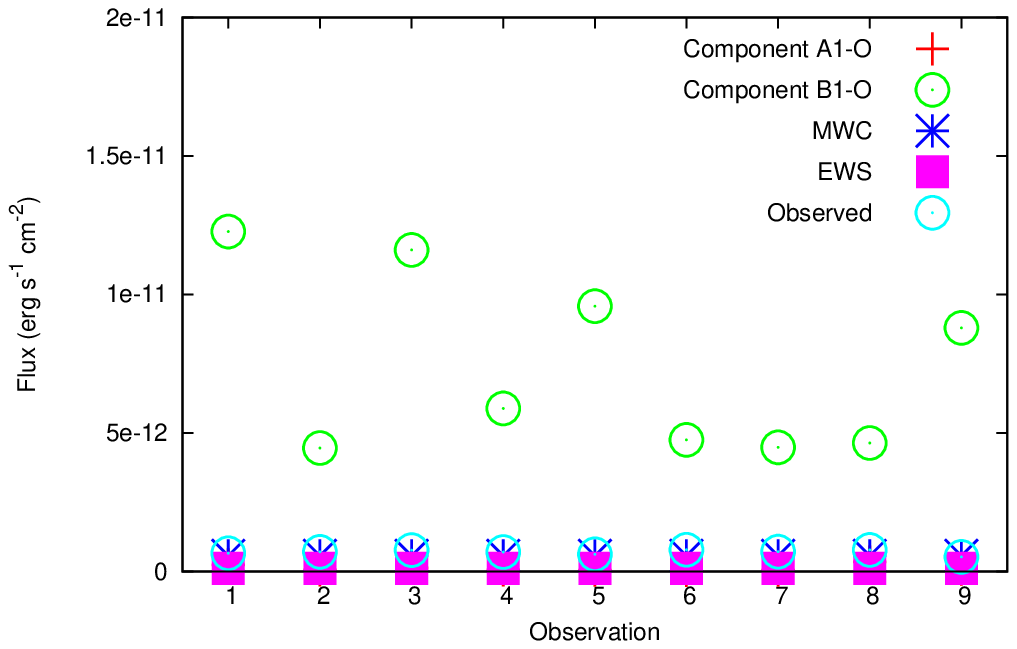}} \\
    \end{tabular}
    \caption{Variation of the intrinsic (top panel) and attenuated
      (bottom panel) X-ray fluxes from wind-wind collisions. The
      fluxes are integrated from the synthetic spectra in the 0.5-8
      keV range. The observed fluxes are also plotted for comparison
      (see Table~\ref{tab:fit5_values}). Clearly, if one sums the
      contributions from all of the emitters in the model the flux
      exceeds the observed values derived from the spectral fits. Note
      the difference in scale between the plots.}
    \label{fig:lx_plots}
  \end{center}
\end{figure}

\begin{figure}
 \begin{center}
   \begin{tabular}{c}
     \resizebox{80mm}{!}{\includegraphics{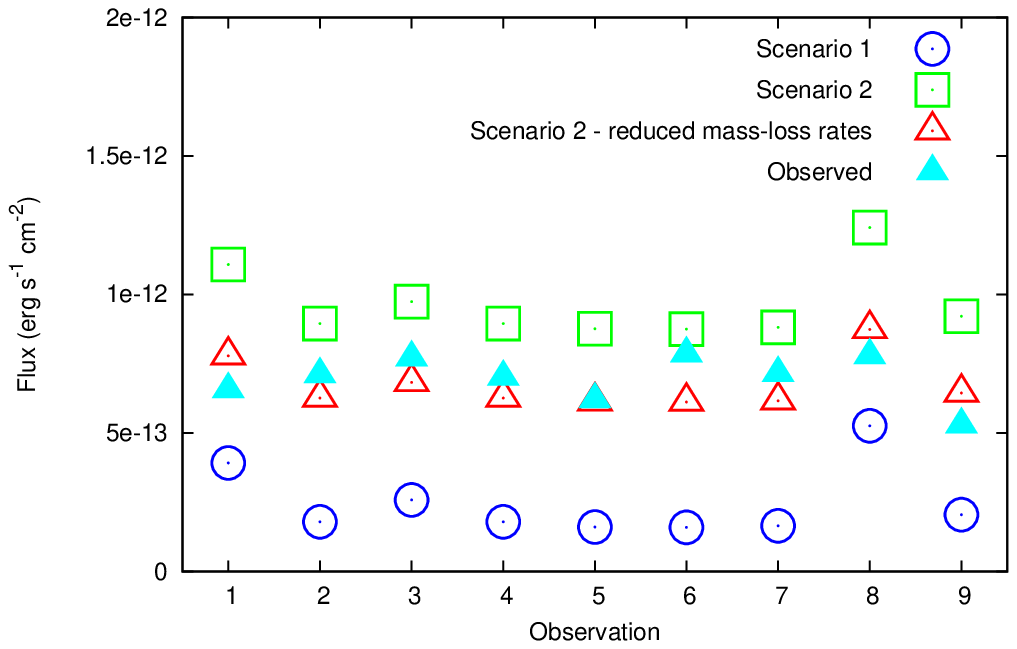}} 
    \end{tabular}
    \caption{Integrated 0.5 - 8 keV fluxes predicted by the model
      compared against the values attained from the spectral
      fits.}
    \label{fig:lx_scenarios}
  \end{center}
\end{figure}

\subsection{Comparison to observations}
\label{subsec:comparison2obs}

The model predictions and observations are in good agreement for a
number of features. For instance, the X-ray emission from wind
collision shocks can explain the hot plasma component derived from the
spectral fits. Despite this, if all of the emitters in the system are
active the total attenuated flux predicted by the model is
$\sim10-20\;$times higher than the fluxes obtained from the spectral
fits, with the dominant contribution to the model flux coming from the
post-shock gas of component B1-O (see bottom panel of
Fig.~\ref{fig:lx_plots}). However, it is unclear whether the
assumption of a normal wind for component B1-O is justified as
\cite{Leung:1979} suggested that this system is
semi-detached. Numerical models of short period, massive star binary
systems by \cite{Dessart:2003} have shown that even a relatively small
mass transfer rate ($\simeq 5\times10^{-6}\Msolpyr$) is unimpeded by
the winds of the stars and can therefore disrupt the apex of the
wind-wind collision region. To gain a better agreement between the
model and observations it would seem necessary for some mechanism to
kill-off the X-ray emission from system B; mass transfer is a viable
option.

Recalling that a comparison of the spectra from QZ~Car and those of
single and binary stars in Carina showed that on the basis of the
spectral shape and the derived plasma temperatures we cannot directly
infer the presence of wind-wind collision shocks, one may be inclined
to neglect the MWC emission also. Yet, if we only consider the X-ray
emission from the individual stars (hereafter Scenario~1) there is a
deficit between model and observation of a factor of $\sim2-3$ (see
Fig.~\ref{fig:lx_scenarios})\footnote{Our adopted stellar parameters
  (Table~\ref{tab:stellar_parameters}) lead to an integrated $L_{\rm
    bol}$ for QZ~Car which is a factor of $\sim2\;$lower than
  determined by \cite{Povich:2010}. Although increasing $L_{\rm bol}$
  by this factor will cause a corresponding increase in $L_{\rm X}$,
  the shortfall between Scenario~1 and the observed fluxes in
  Fig.~\ref{fig:lx_scenarios} would not be remedied.}. Including the
MWC emission (hereafter Scenario~2) remedies the shortfall, and a good
agreement is attained when stellar mass-loss rates are reduced by a
factor of 1.1. Noting that massive star winds are inhomogeneous
\citep[see ][for a recent review]{Puls:2008}, and that results from
detailed observational studies suggest that previous mass-loss rate
estimates \citep[e.g. ][]{Howarth:1989} require scaling down by
factors of 2-5 \citep{Bouret:2003, Repolust:2004, Markova:2004,
  Fullerton:2006, Moffat:2008, Waldron:2010}, this seems to be a
reasonably modest alteration. Notwithstanding the improved agreement
between the model and observed fluxes, the average column derived from
the spectral fits ($\sim0.2\times10^{22}\;$cm$^{-2}$) is significantly
higher than the value calculated for the MWC. This could be indicating
two things. The first is that in using Eq~\ref{eqn:nh} we may be
underestimating the column density to the shocked gas between systems
A and B, or that a more detailed description of the X-ray emission and
absorption to the individual stars is warranted.

Bearing in mind that the X-ray flux from the MWC $\propto 1/ d_{\rm
  sep~A+B}$ (since the shocked gas is adiabatic), we raise the
question of whether a lower limit can be placed on $d_{\rm sep A+B}$
if we make a slightly larger (but still reasonable) reduction in
mass-loss rates? Proceeding with this approach the observed flux level
can be approximately matched if we reduce $\dot{M}$'s by a factor of 3
and decrease $d_{\rm sep A+B}$ to $6\times10^{13}\;$cm, i.e. a factor
of 10 smaller than the upper limit given by \cite{Nelan:2004}. This
alternation in fact causes a negligible change to the column density
calculated for the MWC. However, due to the uncertainty in the orbital
eccentricity of system AB this is a somewhat tentative lower limit.

\section{Discussion}
\label{sec:discussion}

The presence of a hot plasma component with $kT\simeq 2\;$keV in the
spectral fits could be providing evidence for wind-wind collision
shocks in QZ~Car. However, to prevent our model predictions from
considerably overestimating the observed flux we must suppress the
prominent X-ray emission from system B. This is an interesting result
as mass transfer in system B could provide an effective mechanism to
disrupt the wind-wind collision region, and therefore our results
support the previous suggestion by \cite{Leung:1979} of mass transfer
in system B \citep[see also][]{Morrison:1980}. Reassuringly, this
result is unaffected by our adopted distance to QZ~Car - we adopt a
distance of 2.3 kpc which differs from that of \cite{Southworth:2007}
who quote a value of 2.8 kpc. The net effect of using this slightly
larger distance would be a reduction in the calculated fluxes by a
factor of $\sim 0.67$, which would not affect our qualitative
conclusions. We must note, however, that although additional emission
from the MWC is required in our model, a more detailed description of
the emission from the single star than adopted in this work may render
this unnecessary. Furthermore, the ISM column density provides the
dominant absorption to the MWC and the individual stars, therefore, a
small increase/decrease in the ISM column could have implications for
our model results.

This semi-analytical model has nevertheless provided a great deal of
insight. Further progress will require detailed hydrodynamical
modelling which should consider the following factors:
\begin{itemize}
\item The stellar separations in the binary systems are relatively
  small and therefore the interaction between the stellar radiation
  fields may affect the wind acceleration \citep[e.g. inhibition or
    braking, ][]{Stevens:1994, Gayley:1997} which would alter the
  resulting X-ray flux \citep[e.g. ][]{Parkin:2009}.
\item Post-shock gas is in reality multi-temperature and more accurate
  comparisons against the observed spectra will require this to be
  taken into account.
\item The nature of system~B must be properly considered.
\item We account for the radiative behaviour of the shocked gas
  through the $1/\chi$ scaling in Eq.~\ref{eqn:lx}. However, the
  effect of radiative cooling on the dynamics and the X-ray emission
  is likely to be more complicated \citep[e.g.][]{Stevens:1992,
    Myasnikov:1998, Antokhin:2004, Pittard:2009, Parkin:2010,
    Parkinetal:2011}.
\item Contrasting views exist as to the X-ray generation mechanism for
  single O type stars. For instance, recent high resolution analysis
  of O type stars has given evidence for a decrease in X-ray
  temperature in the stellar wind as one tends to larger radii
  \citep{Waldron:2007}. This poses questions for the classic picture
  of X-ray generation by instability driven shocks, whereby higher
  X-ray temperatures are attained at larger radii at which point the
  flow has been accelerated somewhat \citep[][]{Owocki:1988}. Thus, it
  would be appropriate to assess these models in future work, in
  particular examining the spatial and energy dependence of the X-ray
  emission and absorption \citep[e.g.][]{Leutenegger:2010}.

\end{itemize}

\section{Conclusions}
\label{sec:conclusions}

We have presented a series of nine observations of the multiple star
system QZ~Car obtained with {\it Chandra} over a period of roughly 2
years. The spectral fits are characterised by cool, moderate, and hot
temperature plasma components at $kT\simeq 0.2, 0.7,$ and 2 keV,
respectively, a circumstellar absorption of $\simeq
0.2\times10^{22}\;$cm$^{-2}$, and an average flux of
$\simeq7\times10^{-13}\;$erg~s$^{-1}$~cm$^{-2}$. There appears to be
no clear correlation between the fluxes and the orbits of the
constituent binaries. The most compelling evidence for any correlation
is between the high temperature thermal plasma component and the orbit
of the O9.7~I + b2~v binary (system A), although due to limited
statistics the high temperature plasma component is poorly
constrained. Curiously, there is also a deficit between the X-ray flux
expected from the single stars and that derived from the spectral
fits.

A semi-analytical model of QZ~Car was constructed. A stable momentum
balance is not attained between the winds in either the O9.7~I + b2~v
binary (system A) or the O8~III + o9~v binary (system B), and
despite possessing the strongest stellar wind in QZ~Car the O9.7~I
star is a weak emitter (in terms of wind-wind collision emission) due
to the relatively small fraction of its wind being shocked. The higher
fraction of the primary star's wind being shocked in the O8~III +
o9~v binary (system B) makes it the dominant emitter, although the
magnitude of its X-ray emission exceeds the flux level derived from
the spectral fits by more than a factor of 10. The necessity of a
disrupted wind-wind collision region in the O8~III + o9~v binary to
bring the model results and observations into better agreement gives
some compelling evidence in support of \cite{Leung:1979}'s suggestion
of mass transfer.

We conclude that the magnitude and lack of variability in the fluxes
derived from the spectral fits can be well matched by a combination of
X-ray emission from the individual stars and the mutual wind-wind
collision between the two binary systems, albeit with stellar wind
mass-loss rates reduced in-line with the current consensus for
inhomogeneous winds. The observed column density is, however, not well
matched by the model. This may be indicating that a more complex
prescription for the emission from the individual stars is required,
or also that the column density calculation is not completely
appropriate for the mutual wind-wind collision between the two binary
systems. Our analysis places a somewhat tentative lower limit on the
separation of the two binary systems of $\simeq 7\;$AU.

Future analysis would benefit from further observations. A follow-up
X-ray observation with significant enough exposure time to allow a
satisfactory fit with a three temperature plasma model with discrete
absorption components could constrain the column density to the hot
plasma. At radio wavelengths it may in fact be possible to resolve the
separate emission peaks \citep[e.g. ][]{Dougherty:2005}. However, the
sensitivity of current instruments would require a long observation to
attain sufficient statistics, which considering the timescale of the
binary orbits may cause detailed structure to become smeared
\citep[][]{Pittard:2010}. With our results providing support for mass
transfer in the O8~III + o9~v binary, multi-wavelength observations
may reveal a far more complicated picture for QZ~Car
\citep[e.g. $\beta\;$Lyrae; ][]{Ignace:2008}.

\acknowledgements We thank Rodolfo Barb\'{a} and the OWN Team
\citep{Barba:2010} for providing new high-resolution optical
spectroscopic results in advance of publication, Nolan Walborn for
deriving new spectral classifications from those data, as well as for
useful discussions, and the referee for a helpful and insightful
report. ERP was supported in part by a Henry Ellison Scholarship from
the University of Leeds, and by a PRODEX XMM/Integral contract
(Belspo). ERP also thanks Penn State University for their hospitality
during a fruitful visit. JMP gratefully acknowledges funding from the
Royal Society. AFJM is grateful for financial assistance from NSERC
(Canada) and FQRNT (Quebec). YN acknowledges support from the Fonds
National de la Recherche Scientifique (Belgium), the PRODEX XMM and
Integral contracts, and the `Action de Recherche Concert\'{e}e'
(CFWB-Acad\'{e}mie Wallonie Europe). WLW acknowledges partial support
from Chandra grant AR8-9003A. This work is supported by Chandra X-ray
Observatory grant GO8-9131X (PI: L.\ Townsley) and by the ACIS
Instrument Team contract SV4-74018 (PI: G.\ Garmire), issued by the
{\em Chandra} X-ray Center, which is operated by the Smithsonian
Astrophysical Observatory for and on behalf of NASA under contract
NAS8-03060.


\label{lastpage}


\end{document}